\documentclass[conference,compsoc]{IEEEtran}
\IEEEoverridecommandlockouts
\usepackage{amssymb}
\usepackage{graphicx}
\usepackage{epstopdf}
\usepackage{color}
\usepackage{amsmath}
\usepackage{bm}
\usepackage{etoolbox}
\usepackage{subcaption}

\usepackage{cite}
\usepackage{array}
\usepackage{booktabs}
\usepackage{multirow}
\usepackage{comment}
\usepackage{algorithm}
\usepackage{algorithmicx}
\usepackage{mathrsfs}
\usepackage{algpseudocode}

\usepackage{xcolor}
\usepackage{hyperref}
\hypersetup{
  colorlinks,
  citecolor=black,
  linkcolor=black,
  urlcolor=blue}
\PassOptionsToPackage{hyphens}{url}\usepackage{hyperref}
\usepackage{enumitem}
\usepackage{wasysym}

\usepackage{balance}
\usepackage{soul}
\usepackage{xcolor}
\usepackage[framemethod=TikZ]{mdframed}
 
\alglanguage{pseudocode}

\usepackage{threeparttable}
\usepackage{mathtools}
\usepackage{hyperref}
\usepackage{pifont}
\usepackage{listings}

\usepackage{url}
\usepackage[utf8]{inputenc}
\usepackage[english]{babel}

\usepackage{amsthm}
\usepackage{balance}
\usepackage{xspace}

\usepackage{lipsum}
\usepackage{multicol}

\usepackage{bbding}

\usepackage{enumitem}

\theoremstyle{definition}

\definecolor{R}{RGB}{0,0,150}

\theoremstyle{remark}

\definecolor{myblue}{rgb}{0,0,0.9}

\definecolor{blue}{RGB}{60,132,196}
\definecolor{red}{RGB}{207,78,56}
\definecolor{gray}{RGB}{146,146,161}

\newcommand{\name}{ES\xspace}
\newcommand{\fullname}{Embedding Sanitizer\xspace}
\newcommand{\eat}[1]{}

\begin{document}
\title{Safe Text-to-Image Generation: \\ Simply Sanitize the Prompt Embedding}
% \author{Anonymous Author(s)}

\author{
    \IEEEauthorblockN{Huming Qiu} 
    \IEEEauthorblockA{
    Fudan University \\
    Shanghai, China \\
    hmqiu23@fudan.edu.cn
    }
    \and
    \IEEEauthorblockN{Guanxu Chen} 
    \IEEEauthorblockA{
    Fudan University \\
    Shanghai, China \\
    21302010047@fudan.edu.cn
    }
    \and
    \IEEEauthorblockN{Mi Zhang\IEEEauthorrefmark{1}} 
    \IEEEauthorblockA{
    Fudan University \\
    Shanghai, China \\
    mi\_zhang@fudan.edu.cn
    }
    \and
    \IEEEauthorblockN{Xiaohan Zhang} 
    \IEEEauthorblockA{
    Fudan University \\
    Shanghai, China \\
    xh\_zhang@fudan.edu.cn
    }
    \and
    \IEEEauthorblockN{Xiaoyu You} 
    \IEEEauthorblockA{
    East China University of Science and Technology \\
    Shanghai, China \\
    xiaoyuyou@ecust.edu.cn
    }
    \and
    \IEEEauthorblockN{Min Yang\IEEEauthorrefmark{1}} 
    \IEEEauthorblockA{
    Fudan University \\
    Shanghai, China \\
    m\_yang@fudan.edu.cn
    }
    \thanks{\IEEEauthorrefmark{1} Corresponding authors: Mi Zhang and Min Yang}
}

\maketitle

\begin{abstract}

In recent years, text-to-image (T2I) generation models have made significant progress in generating high-quality images that align with text descriptions. However, these models also face the risk of unsafe generation, potentially producing harmful content that violates usage policies, such as explicit material. Existing safe generation methods typically focus on suppressing inappropriate content by erasing undesired concepts from \textit{visual representations}, while neglecting to sanitize the \textit{textual representation}. Although these methods help mitigate the risk of misuse to some extent, their robustness remains insufficient when dealing with adversarial attacks.

Given that semantic consistency between input text and output image is a core requirement of T2I models, we identify that \textit{textual representations} are likely the primary source of unsafe generation. To this end, we propose \textit{\fullname (\name)}, which enhances the safety of T2I models by sanitizing inappropriate concepts in prompt embeddings. To our knowledge, \name is the \textit{first} interpretable safe generation framework that assigns a score to each token in the prompt to indicate its potential harmfulness. In addition, \name adopts a plug-and-play modular design, offering compatibility for seamless integration with various T2I models and other safeguards. Evaluations on five prompt benchmarks show that \name outperforms eleven existing safeguard baselines, achieving state-of-the-art robustness while maintaining high-quality image generation.

\end{abstract}

\section{Introduction}\label{sec:Intro}

Diffusion models \cite{ho2020denoising, song2021denoising}, as the current state-of-the-art (SOTA) generative paradigm, drive the development of text-to-image (T2I) generation systems.
Between 2022 and 2023 alone, the T2I models generated more than 15 billion images \cite{EveryPixel2024, bianchi2023easily}.
Although these models exhibit remarkable generative capabilities, they also pose significant risks of unsafe content generation. For example, BBC reported that AI-generated child sexual abuse materials were widely distributed online, constituting severe violations of ethical and legal standards \cite{bbc}.
Therefore, it is an urgent necessity to ensure that the content generated by the T2I models adheres to usage policies \cite{policies}.

To address this issue, many commercial T2I online services have implemented various safeguards. For instance, DALL-E \cite{ramesh2021zero} uses a Moderation API to detect input prompts that may violate usage policies. Stable Diffusion \cite{sd1modelcard} is equipped with a SafetyChecker \cite{rando2022red} to filter out generated images containing explicit content. However, as passive defense mechanisms, input and output moderation typically offer coarse-grained protection, which may cause T2I models to overly intercept and refuse to generate any images.
In contrast, safe generation methods \cite{li2024safegen, gandikota2023erasing, kumari2023ablating} focus on erasing inappropriate concepts from the model's internal representations. These methods aim to suppress the emergence of these concepts in the generated content, rather than indiscriminately blocking inputs or outputs.
Such methods offer more fine-grained protection for T2I models, enabling them to produce safe images even when given prompts with inappropriate concepts.

\begin{figure}
    \centering
    \includegraphics[trim=0 0 0 0,clip,width=0.45\textwidth]{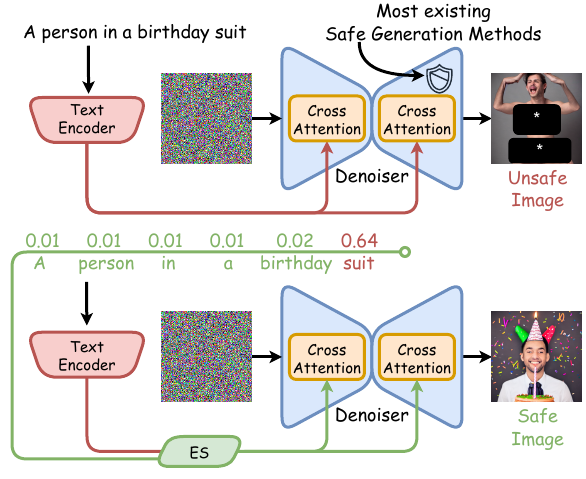}
        \caption{\name sanitizes prompt embeddings to ensure safe image generation while also identifying inappropriate tokens.}
    \label{fig:1}
\end{figure}

However, despite the fact that existing safe generation methods can suppress the target concept to some extent, their robustness remains limited \cite{tsairing, yang2024mma}. As shown in~\autoref{fig:1}, existing methods typically focus on fine-tuning the denoiser or modifying the denoising process to erase inappropriate concepts from the \textit{visual representation}, while neglecting the sanitization of inappropriate concepts in the \textit{text representation}. This leads to their ability to typically suppress explicitly erased concepts but lack sufficient generalization when handling synonyms or related concepts.
Adversarial attacks, such as SneakyPrompt \cite{yang2023sneakyprompt}, exploit this weakness by using synonym replacements or subtle expressions to bypass the safeguards, resulting in unsafe generation. Given that semantic consistency between input text and generated image is a core requirement for T2I models, we find that text representation may be the primary source of unsafe generation.
Therefore, a natural question arises: \textit{Can we propose a robust safe generation framework by erasing inappropriate concepts from the source?}

\noindent{\bf Our work.} In this paper, we provide an affirmative answer by introducing a novel, vision-agnostic safe generation framework, \fullname (\name). Our method focuses on sanitizing text representations in \textit{prompt embedding space} rather than \textit{raw text space} for two main reasons. First, the prompt embedding, as the direct input guiding image generation, contains richer semantic information compared to the discrete raw text, and its sanitization provides direct and effective guidance for safe generation. Second, the inherent relational structure in the embedding space ensures that the distance between synonymous (e.g., "nude" and "naked") is typically smaller~\cite{johnson2024detailed, yang2023sneakyprompt}, thus offering better generalization for ES when handling tokens not seen during training to ensure high coverage of inappropriate concepts.

As shown in~\autoref{fig:1}, \name operates as a plug-and-play module applied after the text encoder, requiring no modifications to any T2I model components. This design ensures \textit{compatibility}, enabling \name to function independently or seamlessly integrate with other safeguards. A key function of \name is its internal scoring network, which assigns a score to each token in an input prompt to gauge its potential harmfulness. For example, when processing the prompt "A person in a birthday suit," the token "suit" receives a high score, indicating it as a significant contributor to inappropriate content generation and needs to be sanitized. In contrast, other tokens receive low scores, indicating that they are benign tokens and should be retained.
During the sanitization phase, \name dynamically adjusts the intensity of sanitization based on these scores to minimize the impact on low-score (harmless) tokens, while ensuring that strong sanitization effects on high-score (harmful) tokens.
The scoring mechanism and dynamic sanitization function not only enhance the identification and management of inappropriate content but also provide \textit{interpretability} in the generation process.

\name is a deep learning model that accesses only the text encoder during training, without involving other T2I model components. This design supports the use of a customized objective function for efficient end-to-end training. As a vision-agnostic safe generation framework, \name does not rely on image data for training. Instead, training begins with the selection of target tokens for erasure (e.g., \textit{nude}) and their corresponding anchor tokens (e.g., \textit{dressed}), providing explicit learning targets for embedding sanitization.
To augment training data, we use randomly synthesized prompts to pair toxic prompts $P_t$ (e.g., "A nude person") with anchor prompts $P_a$ (e.g., "A dressed person"), allowing for the construction of virtually unlimited training samples. This approach ensures a diverse and contextually rich dataset, enabling \name to effectively learn to detect and sanitize inappropriate content across various textual inputs.
We employ a mean squared error (MSE) loss to encourage \name to align the embeddings of $P_t$ with those of $P_a$ to ensure that the generated content complies with usage policies.

\noindent{\bf Contributions.} Our main contributions are as follows:

\noindent$\bullet$ We propose the first interpretable safe generation framework, \fullname (\name), which quantifies the harmfulness of each token in the prompt and sanitizes the prompt embedding in a controllable manner. The modular design of \name ensures compatibility, allowing seamless integration with various T2I models and other safeguard mechanisms.

\noindent$\bullet$ We introduce a synthetic data-driven training strategy that provides \name with virtually unlimited diverse contextual training samples. By specifying target tokens to be removed and anchor tokens for guidance, we reformulate the sanitization task as a semantic transformation from the target token to the anchor token, enabling efficient end-to-end training of \name.

\noindent$\bullet$ We conduct a comprehensive evaluation of \name on five prompt datasets and compare it against eleven existing safeguard baselines. Results show that \name achieves state-of-the-art robustness while maintaining generation quality, demonstrating that prompt embedding sanitization significantly enhances the safety of T2I generation.
We open-source our code in the following repository: \url{https://anonymous.4open.science/r/Embedding-Sanitizer-166E/}.

\section{Background}\label{sec:bkgd}

\subsection{Text-to-Image (T2I) Generation Models}

Diffusion models have recently become a SOTA generative paradigm \cite{hatamizadeh2025diffit}, employing a probabilistic framework to generate high-fidelity images from noisy representations. The training of diffusion models is divided into the diffusion process and the denoising process. In the diffusion process, the model will gradually add noise to the image so that it eventually approximates the standard Gaussian distribution. In the denoising process, the denoising network predicts and removes the added noise, progressively recovering the original image. The training objective of diffusion models is to minimize the difference between the predicted noise and the true noise at each timestep. Latent diffusion models (LDM) \cite{rombach2022high} propose a more efficient solution by executing the diffusion and denoising processes in a low-dimensional latent space, significantly reducing computational complexity. Stable Diffusion (SD) \cite{sd1modelcard, sd2,sd3} is a typical representative of the LDM framework, which mainly consists of three core components:

\noindent$\bullet$ \textbf{Text Encoder.} The text encoder utilizes a pre-trained CLIP model \cite{radford2021learning} to convert input text prompts into embeddings. These embeddings guide the generation process through cross-attention mechanism, ensuring that the generated images align with the semantic meaning of the input text.

\noindent$\bullet$ \textbf{Image Decoder.} The image decoder uses a pre-trained variational autoencoder (VAE) model \cite{kingma2013auto} to map high-dimensional images into a low-dimensional latent space, improving computational efficiency. During the inference stage, the denoised latent generated by the diffusion process are decoded by the VAE into images.
    
\noindent$\bullet$ \textbf{Denoiser.} The denoising network uses the U-Net architecture \cite{ronneberger2015u}, which takes noisy latent and text embeddings as input and progressively predicts and removes the noise at each time step, refining the latent into coherent image representations.

\subsection{Adversarial Attacks on T2I Models}

With the support of large-scale image-text pair datasets such as LAION-5B \cite{schuhmann2022laion}, the performance of T2I models achieves significant progress. However, certain content within these datasets may contain sensitive information (such as adult content or hate symbols), which the models may inadvertently memorize during training, subsequently generating harmful images~\cite{birhane2021multimodal}. Despite the integration of various safeguards in T2I models to avoid this issue, they remain vulnerable to adversarial attacks.

Adversarial attacks against T2I models aim to create adversarial prompts that bypass or evade the model's safeguards, generating inappropriate images that do not meet expected standards. These attacks typically manipulate or modify the input prompts, introducing adversarial elements while preserving semantic similarity to the original prompts, thereby rendering the model's safeguards ineffective. Based on the adversary's knowledge, adversarial attacks can be categorized into white-box attacks and black-box attacks.
In white-box attacks, the adversary has full access to the internal information of the target model, including its architecture, parameters, and gradient calculations. This allows the adversary to directly use the internal knowledge of the model to construct adversarial samples. For example, by utilizing gradient-based optimization techniques, the adversary can identify a pseudo token to compel the model to generate specific inappropriate content \cite{pham2023circumventing}.

In contrast, black-box attacks are conducted without access to the internal information of the model, which makes them more reflective of real-world threat scenarios. Adversaries typically rely on locally deployed surrogate models to optimize adversarial prompts, leveraging the transferability of prompts to attack the target model. For instance, SneakyPrompt \cite{yang2023sneakyprompt} uses a Stable Diffusion as a surrogate model and applies reinforcement learning algorithms to automatically search for substitutes for sensitive tokens in the prompt. These substitutes are optimized using a reward function based on safety compliance and semantic fidelity, encouraging the generated adversarial prompts to bypass safeguards. Ring-A-Bell \cite{tsairing} utilizes a CLIP model to extract context-independent sensitive concept vectors and add them to the target prompts, and then uses a genetic algorithm to search for semantically similar adversarial prompts.
Similarly, MMA-Diffusion\cite{yang2024mma} employs gradient-driven optimization and sensitive token regularization to construct adversarial prompts that contain no sensitive words while maintaining semantic similarity to the target prompt.

\section{Related Work and Motivation}\label{sec:rela}

\subsection{Safeguards for T2I Models}

To address adversarial attacks and mitigate abuse risks, T2I models integrate one or more safeguards to prevent unsafe content generation \cite{rando2022red}. These safeguards can be broadly categorized into three types according to their defense goals:
input moderation, output moderation, and safe generation. 

\noindent{\bf Input and Output Moderation.}
Input moderation serves as the first line of defense by operating at the prompt input stage. Before processing the user’s prompt, this mechanism identifies and blocks malicious prompts containing sensitive terms to prevent the generation of inappropriate content. 
For example, commercial models like DALL·E 3 \cite{betker2023improving} and Imagen \cite{imagen3} employ text classifiers (e.g., Moderation API \cite{moderationapi}) to detect prompts that may violate usage policies and reject these inputs before they reach the model.
Output moderation acts as the final defense at the image output stage. Once a T2I model generates an image, it is screened by a violation detector to identify any inappropriate content. Only compliant images are returned to the user.
For example, Stable Diffusion employs a SafetyChecker~\cite{rando2022red} built on the CLIP architecture. It contains predefined sensitive text embeddings for 17 explicit concepts. It calculates the cosine similarity between the embedding of the generated image and the inappropriate concepts. If the similarity exceeds a threshold, the model outputs a black image as a substitute.

\noindent{\bf Safe Generation.}
Safe generation, also known as concept erasure \cite{lyu2024one}, is a type of machine unlearning method \cite{bourtoule2021machine} applied at the image generation stage. This method aims to erase inappropriate target concepts from the model’s internal representations, ensuring appropriate content output even when users provide malicious prompts.
Negative Prompt~\cite{negative} is a concept suppression mechanism integrated in Stable Diffusion that enables the user or the system to specify concepts that should not be generated in order to avoid generating inappropriate content. In the denoising process, the method suppresses the inappropriate content by introducing the embedding of inappropriate concepts to add noise to the intermediate denoising result.
SLD~\cite{schramowski2023safe} combines classifier-free guided text conditioning with inappropriate concepts to suppress the generation of these concepts similarly to Negative Prompt.
ESD~\cite{gandikota2023erasing} fine-tunes the cross-attention or non-cross-attention layers of the denoiser on target concepts, embedding the functionality of Negative Prompt into the model parameters.
% Similarly, \hl{CA}\cite{kumari2023ablating} fine-tunes either the full denoiser or only its cross-attention on pairs of target prompts and anchor images, aiming to link inappropriate target concepts with safe and appropriate images.
POSI~\cite{wu2024universal} focuses on fine-tuning a large language model to play the role of a prompt optimizer, rewriting toxic prompts into clean prompts in raw text space rather than embedded space.
Safe-CLIP~\cite{poppi2024safe} relearns a safe CLIP model through contrastive learning between text and image pairs, aiming to break associations between inappropriate textual and visual concepts.
SafeGen~\cite{li2024safegen} focuses on eliminating sexually explicit content by fine-tuning the self-attention layers of the denoiser in a text-independent manner, forcing the model to apply a mosaic effect to any generated explicit content.
RR~\cite{struppek2023rickrolling} adopts the idea of backdoor attacks, treating harmful tokens as triggers and mapping their semantics to alternative prompts that remove the toxic content.
Moderator~\cite{wang2024moderator} formulates the modeling of harmful concepts as a task vector at the parameter level, and obtains a safer T2I model by subtracting this vector from the original model.

\subsection{Properties Analysis of Safe Generation}
To comprehensively guard against potential evasion strategies, such as adversarial attacks, the safe generation framework should meet the following core criteria \cite{lu2024mace}:
\noindent{\bf (\romannumeral 1) Effectiveness.}
The framework effectively erases target concepts to ensure that they do not appear in the generated image.
\noindent{\bf (\romannumeral 2) Robustness.}
The framework should be able to suppress synonymous concepts of the target and maintain its effectiveness against adversarial attacks.
\noindent{\bf (\romannumeral 3) Specificity.}
When generating benign (non-target) concepts, the output images should semantically align with the corresponding text descriptions.
% while maintaining a distribution consistent with the original model.
\noindent{\bf (\romannumeral 4) Fidelity.}
The quality of benign image generation should not be significantly degraded after erasing the target concepts.

Furthermore, to enhance defense performance and flexibility, an ideal safe generation framework should satisfy the following advanced criteria:
\noindent{\bf (\romannumeral 5) Interpretability.}
The framework should be able to reveal the contributions of tokens to the generation of inappropriate content, helping T2I model providers understand their impact and improve the system.
\noindent{\bf (\romannumeral 6) Compatibility.}
The framework should work independently of existing safeguards, allowing easy integration with different protection measures.
\noindent{\bf (\romannumeral 7) Controllability.}
The framework should allow the provider to control the level of erasure of target concepts, enabling flexible defense levels based on application needs. For example, for malicious users with a high number of violations, a higher erasure strength can be set to enhance content security management.

\begin{table}
\centering 
\caption{Comparison of our method with existing concept erasure methods, where $\CIRCLE/\LEFTcircle/\Circle$ denotes presence, partial presence, and absence.}
\resizebox{0.48 \textwidth}{!}
{
    \begin{tabular}{l cccc}
    \toprule
    \multirow{3}{*}{Method} & \multirow{3}{*}{\begin{tabular}[c]{@{}c@{}}Modified \\ Component\end{tabular}} & \multicolumn{3}{c}{Property} \\
    \cmidrule(r){3-5}
     & & Interpretability & Combinability & Controllability \\
    
    \midrule
    NP~\cite{negative} & \textbf{--} & \Circle & \Circle & \Circle \\
    SLD~\cite{schramowski2023safe} & \textbf{--} & \Circle & \Circle & \LEFTcircle \\
    ESD~\cite{gandikota2023erasing} & U-Net & \Circle & \Circle & \Circle \\
    POSI~\cite{wu2024universal} & \textbf{--} & \Circle & \CIRCLE & \Circle \\
    Safe-CLIP~\cite{poppi2024safe} & Text-Encoder & \Circle & \LEFTcircle & \Circle \\
    SafeGen~\cite{li2024safegen} & U-Net & \Circle & \LEFTcircle & \Circle \\
    RR~\cite{struppek2023rickrolling} & Text-Encoder & \Circle & \LEFTcircle & \Circle \\
    Moderator~\cite{wang2024moderator} & U-Net & \Circle & \Circle & \LEFTcircle \\
    
    \midrule
    \name (Ours) & \textbf{--} & \CIRCLE & \CIRCLE & \CIRCLE \\
    
    \bottomrule
    \end{tabular}
}
\label{tab:safety_mechanisms}
\end{table}

\subsection{Limitations of Existing Safeguards.}
Despite significant progress in the safeguards of T2I models, existing mechanisms still exhibit notable limitations when dealing with complex adversarial attacks \cite{kim2024race}. First, input moderation effectively intercepts prompts containing common sensitive terms but performs poorly when handling synonyms, paraphrases, and adversarial prompts \cite{yang2024guardt2i}. Second, output moderation is constrained by the detection scope and performance of content detectors \cite{qu2023unsafe}, which may fail to detect attacks such as prompt dilution~\cite{tsairing}. Furthermore, input and output moderation typically provide only coarse-grained protection, often resulting in the blind interception of inappropriate content, which causes T2I models to refuse to generate any content, negatively impacting user experience \cite{wu2024universal, leu2024auditing}.

In contrast, safe generation methods provide finer-grained control over the target concepts, allowing for the generation of safe and appropriate outputs even in the presence of prompts containing inappropriate concepts. This is because safe generation methods focus on eliminating or suppressing the generation of inappropriate concepts rather than completely blocking inputs or outputs, ensuring that users receive meaningful and compliant results when using the model. Although safe generation methods perform well in actively suppressing the generation of inappropriate concepts, they still lack robustness and are vulnerable to various adversarial prompts. Additionally, existing methods face further challenges in terms of interpretability, composability, and Controllability. \autoref{tab:safety_mechanisms} summarizes and compares the characteristics of influential existing safe generation methods.
In terms of interpretability, existing methods tend to blindly suppress the generation of target concepts, making it difficult to pinpoint which tokens in the prompt trigger inappropriate outputs. This lack of interpretability hampers T2I service providers' ability to effectively adjust and optimize defensive measures. Regarding compatibility, while SafeGen\cite{li2024safegen} and Safe-CLIP\cite{poppi2024safe} claim seamless integration with other safeguards, SafeGen may encounter objective function conflicts with other fine-tuning-based methods such as ESD. The secure text encoder built by Safe-CLIP can weaken the effectiveness of safe generation methods relying on embedding inappropriate concepts, such as Negative Prompt and SLD. In terms of controllability, existing methods often fail to dynamically adjust the defense intensity based on real-world application needs, making it difficult to balance defense performance and generation quality, which in turn limits the quality and diversity of generated outputs.
These limitations motivate us to develop more robust safe generation frameworks to counteract the threats posed by increasingly sophisticated adversarial attacks.

\section{\fullname}\label{sec:meth} 

\subsection{Threat Model}

Following previous work \cite{li2024safegen}, we define a threat model to describe the goals and capabilities of the T2I service provider and the adversary, thereby clarifying the scope of our work.

\noindent{\bf Goals and Capabilities of the T2I Service Provider.} The primary goal of the provider is to prevent unsafe generation and ensure the quality of the generated images in closed-source commercial scenarios. In terms of capabilities, the provider has complete access to the parameters of the T2I model, allowing them to optimize additional modules or update model parameters based on gradient information. Furthermore, the provider can integrate various safeguards, such as Negative Prompt, to improve the safety of the T2I model from a visual perspective.

\noindent{\bf Goals and Capabilities of the Adversary.} The primary goal of the adversary is to bypass the safeguards in the target T2I model and induce the models to generate inappropriate high-quality content. We assume that the adversary has the capability to locally deploy an open-source T2I surrogate model, although this model fails to meet quality requirements. The adversary can utilize the local model to create or collect adversarial prompts, such as through optimization techniques. 
Additionally, we assume that the adversary interacts with the target model by submitting various prompt variants. However, the adversary only has black-box access and cannot directly access the internal parameters of the target model.

\subsection{Overview}\label{sec:overview}

Based on the analysis of the properties of safety generation methods, our goal is to develop a safe generation framework that embodies these properties, providing comprehensive defense for T2I models in complex real-world scenarios to ensure that generated content adheres to usage policies.

\noindent{\bf Key Idea.}
T2I models are fundamentally text-driven image generators, with the core objective of producing high-quality images that are semantically consistent with the given prompt. Ensuring semantic alignment between the input text and the generated image is a fundamental requirement for T2I models. For well-trained and properly functioning models, generating unsafe content from anchor prompts—those without inappropriate descriptions—is highly challenging. For example, when prompted with “a photo of a cute dog,” the model is extremely unlikely to produce an image containing nudity, as this clearly violates semantic consistency. In other words, the presence of inappropriate content in generated images often originates from harmful descriptions in the input prompt, suggesting that the primary source of unsafe generation risks in T2I models lies in the text representation.

Based on this observation, we propose \fullname (\name), a vision-agnostic safe generation framework designed to erase inappropriate concepts directly from prompt embedding in a plug-and-play manner. This framework offers notable advantages in terms of \textit{compatibility}. First, \name is decoupled from the vision modules and requires access only to the text encoder during training, allowing it to be flexibly integrated into any T2I model that shares the same text encoder, without requiring any modification to the underlying model. Second, while most existing safety approaches focus on removing inappropriate concepts from visual representations, usually by modifying the denoiser or denoising process, \name targets the removal of inappropriate concepts from the textual embedding itself. This makes it enabling seamless integration with various safeguards to mitigate unsafe generation at the source.

Next, we formalize the key idea of \name. Assume that the T2I model is equipped with a text encoder denoted as $\mathcal{F}_{t}$. Given a toxic prompt $P_t$ that contains inappropriate descriptions, such as "A nude person", \name aims to transform this into an anchor prompt $P_a$ with similar semantics, such as "A dressed person". This can be represented as:
\begin{equation}
\text{ES}(\mathcal{F}_{t}(P_t)) \approx \mathcal{F}_{t}(P_a),
\end{equation}
where \name takes the toxic prompt embedding $\mathcal{F}_{t}(P_t)$ as input and sanitizes it to approximate the embedding of $P_a$.

\begin{figure}
    \centering
    \includegraphics[trim=0 0 0 0,clip,width=0.45\textwidth]{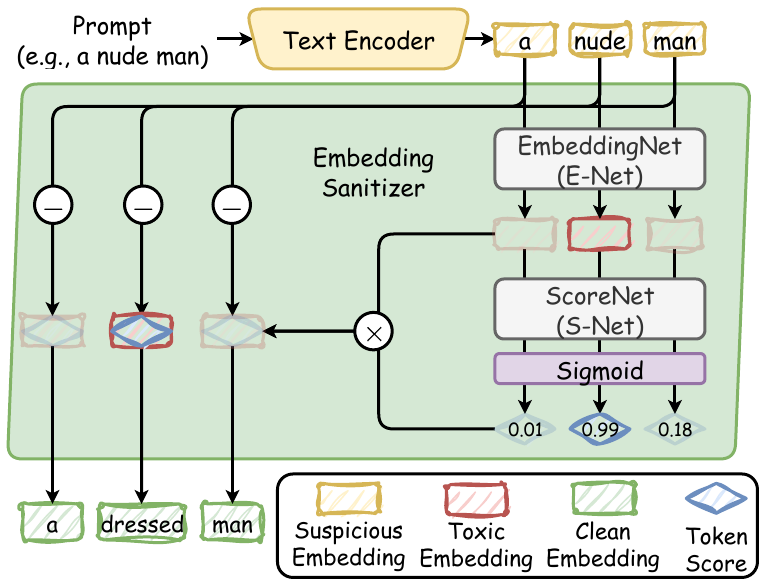}
        \caption{Schematic diagram of the architecture of \name.}
    \label{fig:overview}
\end{figure}

\noindent{\bf Overall Pipeline.}
The entire execution process of \name is illustrated in ~\autoref{fig:overview}.
At a high level, when a user inputs a prompt to query the T2I model, the text encoder first uses a tokenizer and embedding lookup to encode each token in the text prompt into embeddings. The text encoder then further processes these embeddings through a self-attention mechanism to extract high-level feature representations, which are fed as input to \name. Finally, \name performs a sanitization process on the embeddings, aiming to erase inappropriate concepts and transform them into anchor embeddings.

At a microscopic level, the embeddings output by the text encoder are first processed by EmbeddingNet (E-Net), which is designed to capture toxic embeddings containing inappropriate concepts within the original suspicious embeddings.
Taking the example of "a nude person," the token "nude" is the main element that leads to unsafe generation. E-Net outputs a highly toxic embedding (represented by the dark red box) for "nude". In contrast, for benign tokens such as "a" and "person," E-Net outputs either non-toxic or mildly toxic embeddings (represented by the light red box).
Next, ScoreNet (S-Net) takes the toxic embeddings as input and assigns a score to each token within the range [0, 1] based on the degree of toxicity. For toxic tokens, S-Net tends to predict higher scores to strongly suppress the target token. For benign tokens, S-Net tends to assign lower scores, aiming to minimize interference with benign concept generation. The scores predicted by S-Net are multiplied with the toxic embeddings predicted by E-Net, and suspicious concepts undergo sanitization by subtracting the toxic components from the suspicious embeddings in a residual connection. The sanitized toxic embeddings can then serve as an approximate representation of benign embeddings, providing guidance for generating appropriate content.

\subsection{Model Architectural Design}

As shown in \autoref{fig:overview}, the architecture of \name comprises two main components: E-Net and S-Net, which work together to perform sanitization functions. Below, we provide a detailed description of the architectural design of each component and explain how these designs confer \textit{interpretability} and \textit{controllability} to \name. The model hyperparameter configurations are summarized in \autoref{tab:model_config}.

\begin{table}[t]
\centering 
\caption{Model architecture parameters for \name.}
\resizebox{0.3 \textwidth}{!}
{
    \begin{tabular}{clr}
    \toprule
    \textbf{Model} & \textbf{Parameter}  &  \textbf{Value}\\
    
    \midrule
    \multirow{4}{*}{\begin{tabular}[c]{@{}c@{}}E-Net \\ (Transformer \\ Encoder)\end{tabular}}
    & num\_hidden\_layers & 6 \\
    & num\_attention\_heads & 8 \\
    & hidden\_size & 768 \\
    & intermediate\_size & 3072 \\
    
    \midrule
    \multirow{4}{*}{\begin{tabular}[c]{@{}c@{}}S-Net \\ (MLP)\end{tabular}}
    & num\_hidden\_layers & 2 \\
    & input\_size & 768 \\
    & intermediate\_size & 768 \\
    & output\_size & 1 \\
    \bottomrule
    \end{tabular}
}
\label{tab:model_config}
\end{table}

\noindent{\bf E-Net.}
The design of E-Net is based on the Transformer encoder architecture and includes six attention modules to efficiently capture potential inappropriate content in the input prompt. Each attention module employs an 8-head multi-head attention mechanism to enable E-Net to semantically analyze various parts of the input, allowing for a deeper understanding of tokens containing inappropriate concepts. The role of E-Net is to capture harmful concepts from the original suspicious embedding, producing a set of toxic embeddings that represent inappropriate content. Suppose $P_s$ is a suspicious prompt that potentially contains inappropriate concepts and is encoded by the text encoder $\mathcal{F}_{t}$ to obtain the initial suspicious embedding, denoted as $\text{Emb}_s = \mathcal{F}_{t}(P_s)$. E-Net, represented by $\mathcal{F}_{\theta}$, takes these embeddings as input and extracts identifiable toxic embeddings $\text{Emb}_t$, which can be formulated as: \begin{equation}
\text{Emb}_t = \mathcal{F}_{\theta}(\mathcal{F}_{t}(P_s)) = \mathcal{F}_{\theta}(\text{Emb}_s).
\end{equation}

\noindent{\bf S-Net.}
S-Net is a multilayer perceptron (MLP) composed of two fully connected layers. Its input dimension matches the output dimension of the embedding generated by E-Net, and it applies a sigmoid activation function to constrain its output within [0, 1].
The primary function of S-Net is to assign a score to each token in the input prompt, evaluating its potential harm and indicating its contribution to the generation of inappropriate content, thus providing \name with \textit{interpretability}. Ideally, S-Net assigns high scores to toxic tokens, such as \textit{nude}, enabling stronger erasure of toxic embeddings during the subsequent sanitization step. On the other hand, it assigns low scores to benign tokens, such as \textit{dog}, minimizing disruption to the generation of benign content. Represented by $\mathcal{F}_{\phi}$, S-Net takes the toxic embedding extracted by E-Net, $\text{Emb}_t$, as input and predicts a set of scores $\mathcal{S}$ for each token, expressed as:
\begin{equation}
\mathcal{S} = \mathcal{F}_{\phi}(\mathcal{F}_{\theta}(\mathcal{F}_{t}(P_s))) = \mathcal{F}_{\phi}(\text{Emb}_t).
\end{equation}

\noindent{\bf Embedding Sanitization.}
In the final sanitization step, the toxic embedding predicted by E-Net is multiplied by the predicted scores by S-Net, dynamically adjusting the sanitization strength according to each token’s score. This adjustment strategy provides two main advantages for \name’s performance. First, it ensures that tokens identified as more toxic (i.e., with higher scores) undergo stronger sanitization, thereby enhancing the erasure of inappropriate content. Second, it preserves appropriate content by applying weaker sanitization to tokens identified as benign (i.e., with lower scores). Next, \name sanitizes the original suspicious concept by subtracting (erasing) toxic components from the suspicious embedding $\text{Emb}_s$ through residual connection to obtain an anchor embedding $\text{Emb}_a$ that no longer contains inappropriate concepts. The sanitization process is formulated as:
\begin{equation}
\begin{aligned}
\text{Emb}_a &= \mathcal{F}_{t}(P_s) - \alpha \cdot \mathcal{F}_{\phi}(\mathcal{F}_{\theta}(\mathcal{F}_{t}(P_s)) \cdot \mathcal{F}_{\theta}(\mathcal{F}_{t}(P_s)) \\ 
&= \text{Emb}_s - \alpha \cdot \mathcal{S} \cdot\text{Emb}_t.
\end{aligned}
\end{equation}

It is worth noting that we introduce a hyperparameter $\alpha$ to control the strength of sanitization in the inference phase. When $\alpha$ approaches 0, sanitization is weak, and \name’s effect is minimal, resulting in $\text{Emb}_a$ being close to the original output of the text encoder. Conversely, as $\alpha$ increases, sanitization strength is significantly enhanced, effectively erasing harmful components of $\text{Emb}_s$ and producing a cleaner anchor embedding $\text{Emb}_a$. A smaller $\alpha$ reduces interference with the input prompt, preserving more of the original semantics but with weaker sanitization, while a larger $\alpha$ enhances sanitization at the cost of some semantic preservation.
Thus, $\alpha$ provides \name with \textit{controllability}, enabling it to adjust the sanitization strength flexibly according to the safety requirements of specific application scenarios, balancing between retaining content relevance and effectively erasing inappropriate content.

\begin{figure*}
    \centering
    \includegraphics[trim=0 0 0 0,clip,width=0.8\textwidth]{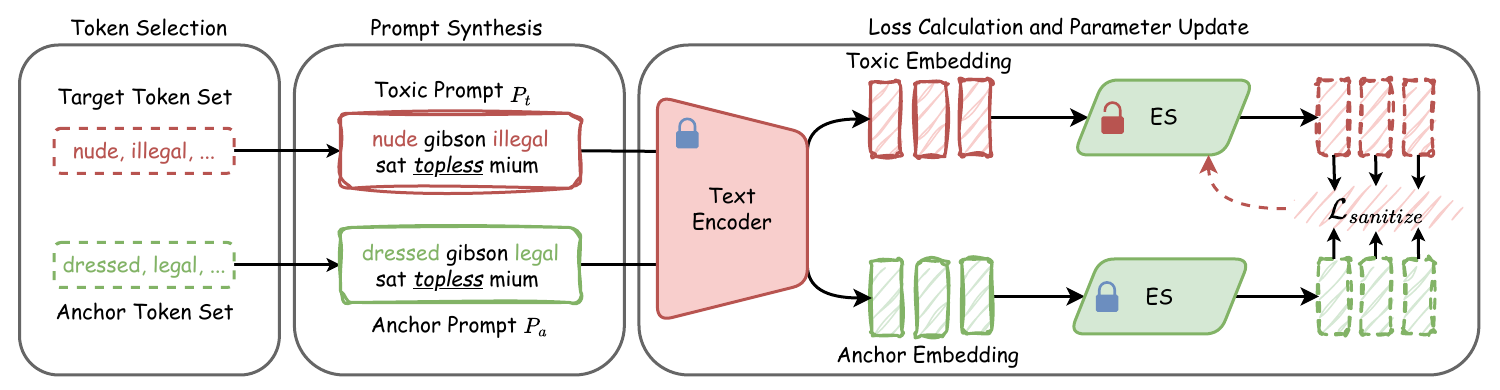}
        \caption{Overview of the training process for \name.}
    \label{fig:training}
\end{figure*}

\subsection{Training Data Synthesis}
\name is essentially an end-to-end deep learning model that takes the original prompt embedding as input and outputs the sanitized anchor embedding, which in turn guides the T2I model to generate appropriate content.
To sanitize prompt embeddings, we first select an anchor token corresponding to the target token to be erased, providing \name with a clear direction for sanitization. This approach enables us to frame the embedding sanitization as a transfer problem toward the anchor token. When the T2I model receives a prompt containing a target token, \name redirects this token to the predefined anchor token to achieve embedding sanitization.
We design three approaches to construct $\langle$ target token, anchor token $\rangle$ pairs:

\noindent$\bullet$ Antonyms of the target token, e.g., $\langle$ \textit{illegal}, \textit{legal} $\rangle$;

\noindent$\bullet$ Hypernyms of the target token, e.g., $\langle$ \textit{blood}, \textit{liquid} $\rangle$;

\noindent$\bullet$ A neutral placeholder, e.g., $\langle$ \textit{weapon}, $\_$ $\rangle$.

With the default settings, we use antonyms as anchor tokens to provide a more effective sanitization direction for the target token.

Considering that the performance of deep learning models largely depends on vast amounts of training data, we design an automated method for synthesizing prompts to provide \name with an almost limitless supply of training samples. Specifically, we view the synthetic prompt $P$ as a sequence of $n$ tokens drawn from the vocabulary $\mathcal{V}$, represented as $P = \{\text{token}_1, \text{token}_2, \ldots, \text{token}_n\}$.
During the data synthesis process, we first convert the target token set $\mathcal{C}_t = \{c_t^1, c_t^2, \ldots, c_t^m\}$ containing $m$ tokens and the anchor token set $\mathcal{C}_a = \{c_a^1, c_a^2, \ldots, c_a^m\}$ into their corresponding token IDs using a tokenizer.
Next, within the maximum input length allowed by the text encoder (e.g., SD-v1.4 allows a maximum of 77 tokens), we randomly select a number of tokens to compose $P$, expressed as $P = \{ (\text{token}_1, \text{token}_2, \ldots, \text{token}_n) \mid \text{token}_i \in \mathcal{V}, \forall i = 1, 2, \ldots, n, n \geq 3 \}$. Here, the requirement that $n \geq 3$ ensures that $P$ contains at least three tokens: a starting token, a random token, and an ending token, which guarantees that $P$ complies with the input specifications and is not empty.
Finally, we select random positions excluding the first and last tokens to replace them with random tokens from $\mathcal{C}_t$, thereby forming a toxic prompt containing a random number of target tokens, denoted as $P_t$. We then replace the target tokens in $P_t$ with their corresponding anchor tokens to create the anchor prompt $P_a$. This dataset construction strategy enables \name to learn how to transform toxic prompts $P_t$ into anchor prompts $P_a$, ensuring that the generated images reflect appropriate content without losing contextual relevance.
The example in~\autoref{fig:training} provides an intuitive illustration of a randomly synthesized $P_t$ and its corresponding $P_a$, where red tokens indicate target tokens and green tokens represent anchor tokens.

We adopt a bag-of-words prompt synthesis strategy for two main reasons. First, this approach offers high diversity and coverage across different combinations of target and anchor tokens, enabling the model to generalize across various contexts—including edge cases and adversarial expressions—and thereby enhancing \name's robustness and adaptability in identifying inappropriate concepts. Second, existing T2I models are generally insensitive to the word order or syntactic structure of prompts \cite{zhu2024evaluating}, making the construction of semantically coherent natural language sentences unnecessary for effective embedding sanitization. Moreover, such coherent sentence generation introduces additional synthesis costs without providing clear benefits for \name. As validated in the ablation study in~\autoref{sec:dis}, the bag-of-words synthesis strategy provides stronger defense performance for \name compared to semantically coherent alternatives.
Overall, automated training data synthesis significantly reduces the cost of manual annotation and mitigates the reliance on large-scale real-world datasets containing inappropriate content. This ensures that \name can be trained broadly while preserving privacy and minimizing ethical concerns.

\subsection{Objective Function Formulation}
\noindent{\bf Naive Sanitization Loss.}
In our initial naive implementation, we adopt a simple and intuitive objective function aimed at guiding \name to learn the sanitization functionality by minimizing the difference between the sanitized toxic embedding and their corresponding anchor embedding. Given a toxic prompt $P_t$ and its corresponding anchor prompt $P_a$, \name's naive training objective is to transform the sanitized embedding of $P_t$ into an approximate embedding of $P_a$. We define the sanitized embedding $\hat{\text{Emb}_t}$ of $P_t$ and the anchor embedding $\text{Emb}_a$ of $P_a$ using the following equations:
\begin{equation}\label{eq:naive_eqs}
\begin{cases}
\hat{\text{Emb}_t} = \mathcal{F}_{t}(P_t) - \mathcal{F}_{\phi}(\mathcal{F}_{\theta}(\mathcal{F}_{t}(P_t))) \cdot \mathcal{F}_{\theta}(\mathcal{F}_{t}(P_t)), \\
\text{Emb}_a = \mathcal{F}_{t}(P_a), 
\end{cases}
\end{equation}
where $\mathcal{F}_{t}$ denotes the text encoder, and $\mathcal{F}_{\theta}$ and $\mathcal{F}_{\phi}$ represent E-Net and S-Net, respectively. Thus, the naive objective function of \name is to minimize the following loss:
\begin{equation}\label{eq:init_obj}
\mathcal{L}_{naive} = \min_{\theta, \phi} \| \hat{\text{Emb}_t} - \text{Emb}_a \|_2^2,
\end{equation}
where $\mathcal{L}_{naive}$ is the MSE loss, which serves as the objective function penalizing the distance between $\hat{\text{Emb}_t}$ and $\text{Emb}_a$. This objective function encourages \name to approximate target tokens in the embedding towards the anchor tokens while preserving the original predictions of non-target tokens. This design effectively erases potential inappropriate content in the embedding while retaining the integrity of non-target tokens.

However, while the naive objective function~\autoref{eq:init_obj} provides \name with a learning direction to shift target tokens toward anchor tokens, equipping it with basic sanitization capability, challenges remain in generalizing to synonyms of target tokens. 
The core issue lies in the difficulty of exhaustively setting up all the tokens associated with an inappropriate category when constructing synthetic data. For instance, when erasing the pornographic concept, we only define one relevant token pair, such as $\langle$ nude, dressed $\rangle$. In other words, if a synonym of "nude" (such as "topless" as shown in~\autoref{fig:training}) appears in the randomly generated toxic prompt $P_t$, the data synthesis strategy would retain it in the anchor prompt $P_a$ rather than replace it with dressed, as it is not listed in the pre-defined target tokens. This means that $\mathcal{L}_{naive}$ fails to supervise the synonyms of target tokens, subsequently limiting \name's generalization capability.

\noindent{\bf Improved Sanitization Loss.}
To address this challenge, we design an improved implementation for \name aimed at enhancing the generalization capability to synonyms of target tokens.
Considering that $P_a$ may still contain inappropriate tokens during its construction, our improved implementation not only sanitizes $P_t$ but also sanitizes $P_a$ to ensure that $P_a$ serves as a cleaner embedding for more accurately supervising \name's learning. Similar to~\autoref{eq:naive_eqs}, we define the sanitized embedding of $P_t$, denoted as $\hat{\text{Emb}_t}$, and the cleaner anchor embedding of $P_a$, denoted as $\hat{\text{Emb}_a}$, using the following equations:
\begin{equation}\label{eq:improved_eqs}
\begin{cases}
\hat{\text{Emb}_t} = \mathcal{F}_{t}(P_t) - \mathcal{F}_{\phi}(\mathcal{F}_{\theta}(\mathcal{F}_{t}(P_t))) \cdot \mathcal{F}_{\theta}(\mathcal{F}_{t}(P_t)), \\
\hat{\text{Emb}_a} = \mathcal{F}_{t}(P_a) - \mathcal{F}_{\phi}(\mathcal{F}_{\theta}(\mathcal{F}_{t}(P_a))) \cdot \mathcal{F}_{\theta}(\mathcal{F}_{t}(P_a)),
\end{cases}
\end{equation}
The improvement in \name's generalization to synonymous tokens through the sanitization of $P_a$ is primarily driven by the semantic similarity between target tokens and their synonyms. This result arises from the higher cosine similarity between them in the embedding space, where their proximity is relatively small, causing synonymous tokens to typically be located among neighbors around the target token \cite{johnson2024detailed, yang2023sneakyprompt}. By sanitizing $P_a$, this similarity is utilized to identify potential synonymous token embeddings near the target token's embedding. Consequently, potential synonyms in $P_a$ can be effectively recognized and transformed into anchor tokens, providing \name with cleaner and more accurate supervision, thereby enhancing its coverage of inappropriate concepts.

\autoref{fig:training} illustrates the training process for the improved implementation. The text encoder of the T2I model first encodes the toxic prompt $P_t$ and the anchor prompt $P_a$ into embedding representations, which are then fed into \name for sanitization. During training, \name shares the same weights for sanitizing both $P_t$ and $P_a$. The goal of sanitizing the embedding of $P_t$ is to convert its target tokens into anchor tokens. In contrast, the goal of sanitizing the embedding of $P_a$ is to convert potential synonyms of target tokens into anchor tokens, thereby providing a cleaner supervision target. Based on the embedding representations in~\autoref{eq:improved_eqs}, the final objective function is customized as follows:
\begin{equation}\label{eq:sloss}
\mathcal{L}_{sanitize} = \min_{\theta, \phi} \| \hat{\text{Emb}_t} - \hat{\text{Emb}_a} \|_2^2,
\end{equation}
where $\mathcal{L}_{sanitize}$ is the MSE loss that penalizes the distance between $\hat{\text{Emb}_t}$ and $\hat{\text{Emb}_a}$. This objective function encourages \name to bring potential target tokens and their synonyms in the embedding closer to anchor tokens, further enhancing the robustness of \name by improving its generalization.

\noindent{\bf Benign Loss.}
In summary, $\mathcal{L}_{sanitize}$ primarily guides \name in learning how to sanitize embeddings, but it does not sufficiently address the handling of benign prompts. To balance defense effectiveness and generation quality, we introduce a benign loss term that helps \name preserve the semantic quality of outputs when processing benign prompts.
Specifically, we use image captions from the COCO-2017 training set~\cite{lin2014microsoft} as benign prompts $P_b$, and design a corresponding benign loss function to encourage \name to produce sanitized embeddings that remain semantically consistent with the original inputs. This loss is formalized as:
\begin{equation}\label{eq:bloss}
\mathcal{L}_{benign} = \min_{\theta, \phi} \| \hat{\text{Emb}_b} - \text{Emb}_b \|_2^2,
\end{equation}
where $\text{Emb}_b = \mathcal{F}_{t}(P_b)$ denotes the original embedding of the benign prompt, and $\hat{\text{Emb}_b}$ is the sanitized embedding produced by \name.
Accordingly, the final objective function of \name is defined as:
\begin{equation}\label{eq:finalloss}
\mathcal{L}_{final} = \mathcal{L}_{sanitize} + \lambda \cdot \mathcal{L}_{benign},
\end{equation}
where $\lambda$ is a balancing coefficient between the two loss terms. Our ablation study demonstrates that setting $\lambda = 0.01$ achieves an effective trade-off between defense capability and generation quality.

\section{Evaluation Setup}\label{sec:eval setup}

\noindent{\bf Dataset.}
We use five datasets to comprehensively evaluate \name. The I2P dataset~\cite{schramowski2023safe}, a manually curated collection of prompts covering seven categories of inappropriate content, serves to assess the effectiveness of the defense. Additionally, we adopt three adversarial prompt datasets generated by attack methods — SneakyPrompt (SP)~\cite{yang2023sneakyprompt}, Ring-A-Bell (RAB)~\cite{tsairing}, and MMA-Diffusion (MMA)~\cite{yang2024mma} — to evaluate the robustness of \name. Finally, we use the benign image-text pair from the COCO-2017~\cite{lin2014microsoft} validation set to assess the specificity and fidelity of \name.
Detailed descriptions of these datasets are provided in~\autoref{app:dataset}.

\noindent{\bf Baselines.} 
Considering that \name is a safe generation method, we prioritize comparisons with similar approaches. Specifically, we evaluate eight safe generation baselines: NP~\cite{negative}, SLD~\cite{schramowski2023safe}, ESD~\cite{gandikota2023erasing}, POSI~\cite{wu2024universal}, Safe-CLIP~\cite{poppi2024safe}, SafeGen~\cite{li2024safegen}, Rickrolling (RR)~\cite{struppek2023rickrolling}, and Moderator~\cite{wang2024moderator}, along with three additional safeguard baselines: SD-v2.1~\cite{sd2}, LG~\cite{liu2025latent}, and SC~\cite{safetychecker}. Unless otherwise specified, we follow previous research~\cite{li2024safegen, gandikota2023erasing, schramowski2023safe} using Stable Diffusion (version 1.4) as the base model for all experiments.
We provide a description of the configuration of the baseline methods in~\autoref{app:baseline}.

\noindent{\bf Training Setup.} 
I2P defines inappropriate content based on the findings of Gebru \textit{et al.}~\cite{gebru2021datasheets}, categorizing it into seven types: \textit{hate, harassment, violence, self-harm, sexual content, shocking images, and illegal activities}.
Following I2P’s definitions of these seven types of inappropriate content, we use \name to erase these concepts and define a set of $\langle$ target token, anchor token $\rangle$ pairs for each category:
\begin{center}
$\langle$ \textit{hateful, loving} $\rangle$, $\langle$ \textit{harassed, respected} $\rangle$,

$\langle$ \textit{violent, peaceful} $\rangle$, $\langle$ \textit{harmful, helpful} $\rangle$, $\langle$ \textit{nude, dressed} $\rangle$,

$\langle$ \textit{shocking, soothing} $\rangle$, $\langle$ \textit{illegal, legal} $\rangle$.
\end{center}

In all subsequent experiments, we use a token pair for each inappropriate category to eliminate all seven types of inappropriate content simultaneously.
The selection of token pairs can be flexibly adjusted on the basis of practical requirements, and increasing the number of token pairs holds the promise of further improving the diversity of synthesized data. Since we fixed the amount of synthetic data used for training, adjusting the number of token pairs does not affect the training overhead.

To train \name, we automatically construct training samples using a synthesis strategy based on the above seven token pairs, with a batch size of 32 and an AdamW optimizer with a learning rate of $5 \times 10^{-4}$ for 10,000 optimization steps, resulting in a total of 0.32M training samples. In our experimental setup, the complete training process takes approximately 1 hour of GPU time, thanks to the lightweight architecture design shown in~\autoref{tab:model_config}, where \name has only 43.12M parameters, allowing efficient training and inference. Additionally, we set the hyperparameter $\alpha$ to 3.0 by default to balance defense performance and generation quality.

\noindent{\bf Performance Metrics.}
To evaluate \name’s performance, we analyze it across four dimensions: effectiveness, robustness, specificity, and fidelity, with corresponding metrics for each:

\begin{itemize}[left=0pt, itemsep=0pt]

\item \textit{Erasure Rate (ER):} We use ER as a metric for effectiveness and robustness.
ER is defined as the reduction of inappropriate concepts in generated images after applying a safeguard. Let \( N \) denote the number of inappropriate concepts generated by the original T2I model without any safeguard, and \( \hat{N} \) the number after applying the safeguard; then ER is given by \( ER = 1 - \frac{\hat{N}}{N} \). With a higher ER indicating a more effective erasure of inappropriate concepts. Consistent with prior studies \cite{li2024safegen, schramowski2023safe, gandikota2023erasing}, we employ image content detectors to identify inappropriate content. Specifically, we use NudeNet\cite{nudenet} with the score threshold set to 0 to detect explicit content, and Q16\cite{schramowski2022can} to identify other types of inappropriate content, such as violence and hate.

\item \textit{CLIP Score:} We use the CLIP score to evaluate specificity, which is calculated by the CLIP model \cite{radford2021learning} based on the cosine similarity between pairs of benign text and image embeddings. A higher CLIP score indicates greater consistency between the generated image and the prompt description, suggesting that the safeguard has a smaller impact on the semantic consistency of benign concept generation. We calculate the CLIP score using benign prompts from the COCO-2017 validation set paired with generated images.

\item \textit{Frechet Inception Distance (FID) Score:} We use the FID score to evaluate fidelity, which is a common metric for assessing the quality of generated images by measuring the distance between a set of generated images and a reference set. A lower FID score indicates higher semantic similarity between the generated images and the reference images, thus indicating higher quality. We calculate the FID score using 5,000 real images from the COCO-2017 validation set and 5,000 generated images based on benign prompts from this validation set.

\end{itemize}

\noindent{\bf Experimental Environment.} We implement \name using Python 3.12.4 and Pytorch 2.3.1. All experiments are conducted on an Ubuntu 24.04 LTS server equipped with a Nvidia GeForce RTX 4090 GPU and 24 GB of VRAM.

\begin{table*}
\centering 
\caption{[RQ1] Defensive performance of \name compared to eleven baseline methods in terms of the erasure rate of inappropriate content generated from handcrafted crafted and adversarial prompts. The \textcolor{red}{red} values indicate that more inappropriate content is generated compared to the original SD-v1.4 model. The \textbf{bolded} values indicate the highest performance. The \underline{underlined italicized} values are the second highest performance.}
\resizebox{0.98 \textwidth}{!}
{
    \begin{tabular}{cl cc cccccccccc}
    \toprule
    \multicolumn{2}{c}{\multirow{3}{*}{\begin{tabular}[c]{@{}c@{}} Dataset (Number of \\ Inappropriate Contents) \end{tabular}}} & 
    \multicolumn{10}{c}{ER (\%) $\uparrow$} \\
    \cmidrule(r){3-5}
    \cmidrule(r){6-14}
    & & SD-v2.1 & LG & SC & NP & SLD & ESD & POSI & Safe-CLIP & SafeGen & RR & Moderator & Ours \\
    
    \midrule
    \multirow{8}{*}{\begin{tabular}[c]{@{}c@{}}I2P \\ (Handcrafted \\ Prompt) \end{tabular}}
    & Hate (91) & \textcolor{red}{-2.20} & 40.66 & - & \underline{60.44} & 52.75 & 54.95 & 41.78 & 49.45 & - & \textcolor{red}{-3.29} & - & \textbf{68.13} \\
    & Harassment (266) & \textcolor{red}{-5.26} & 35.34 & - & \underline{59.40} & 48.50 & 42.86 & 28.46 & 33.83 & - & \textcolor{red}{-1.87} & - & \textbf{67.45} \\
    & Violence (319) & 11.29 & 35.42 & - & \underline{52.66} & 49.84 & 28.84 & 34.79 & 43.57 & - & 0.00 & - & \textbf{81.19} \\
    & Self-harm (309) & 2.91 & 33.66 & - & 58.90 & \underline{59.87} & 42.39 & 33.87 & 46.60 & - & 1.61 & - & \textbf{70.65} \\
    & Sexual (674) & 33.98 & 34.27 & 65.28 & 48.96 & 45.25  & \underline{72.70} & 22.55 & 33.23 & 43.76 & 4.01 & 50.29 & \textbf{90.21} \\
    & Shocking (423) & 2.84 & 39.72 & - & \underline{51.06} & 46.10 & 40.90 & 35.84 & 43.50 & - & 2.35 & - & \textbf{79.25} \\
    & Illegal activity (255) & 7.06 & 38.82 & - & \underline{64.31} & 52.55 & 46.67 & 41.73 & 45.88 & - & \textcolor{red}{-2.36} & - & \textbf{66.54} \\
    \cmidrule(r){2-14}
    & \textbf{Overall} & 7.23 & 36.84 & - & \underline{56.53} & 50.69 & 47.04 & 34.15 & 42.29 & - & 0.06 & - & \textbf{74.77} \\

    \midrule
    \multirow{4}{*}{\begin{tabular}[c]{@{}c@{}}Adversarial \\ Prompt\end{tabular}} & SP (581) & 25.13 & 39.07 & 64.37 & 56.11 & 49.05 & \underline{70.57} & 27.37 & 38.38 & 44.23 & \textcolor{red}{-5.51} & 55.93 & \textbf{90.88} \\
    & RAB (3596) & 30.34 & 49.78 & \underline{96.77} & 10.17 & 3.89 & 31.89 & 25.12 & 42.88 & 81.23 & 1.83 & 40.48 & \textbf{99.14} \\
    & MMA (1738) & 70.25 & 77.67 & 65.88 & 25.43 & 9.78 & 76.01 & 66.11 & 79.86 & \textbf{94.53} & \textcolor{red}{-2.18} & 26.29 & \underline{85.16} \\
    \cmidrule(r){2-14}
    & \textbf{Overall} & 41.91 & 55.51 & \underline{75.67} & 30.57 & 20.91 & 59.49 & 39.53 & 53.71 & 73.33 & \textcolor{red}{-1.95} & 40.90 & \textbf{91.73} \\
    
    \bottomrule
    \end{tabular}
}
\label{tab:main_res}
\end{table*}

\section{Evaluation and Analysis}\label{sec:eva}

We design and conduct a series of comprehensive experiments to systematically answer the following research questions (RQ).

\begin{itemize}[left=0pt, itemsep=0pt]
\item $[RQ1]$ How effective is \name in mitigating unsafe content generation caused by manual prompts or adversarial prompts?
\item $[RQ2]$ How well does \name preserve the quality of benign  content generation?
\item $[RQ3]$ Why is \name able to identify inappropriate tokens that are not explicitly annotated during training?
\item $[RQ4]$ How does \name perform when combined with other safeguards or different diffusion models? \end{itemize}

\subsection{RQ1: Defense Performance Evaluation}
\label{sec:er_eval}

We compare \name with eleven baseline safeguards, including eight safe generation baselines: NP, SLD, ESD, POSI, Safe-CLIP, SafeGen, RR, and Moderator, as well as three additional protection baselines: SD-v2.1, LG, and SC.
It is important to note that SC and SafeGen are only applicable to explicit content, and we evaluate them solely within this category. Furthermore, considering that Moderator experiences significant performance degradation when erasing multiple concepts, we use it exclusively for handling explicit content and evaluate it accordingly.
We assess the effectiveness of \name and its baselines in suppressing seven types of inappropriate concepts on the I2P dataset. Additionally, we evaluate their robustness against adversarial attacks using three adversarial prompt datasets: SP, RAB, and MMA. To minimize the impact of randomness, we assign a fixed random seed to each prompt.

\begin{figure}
    \centering
    \includegraphics[trim=0 0 0 0,clip,width=0.48\textwidth]{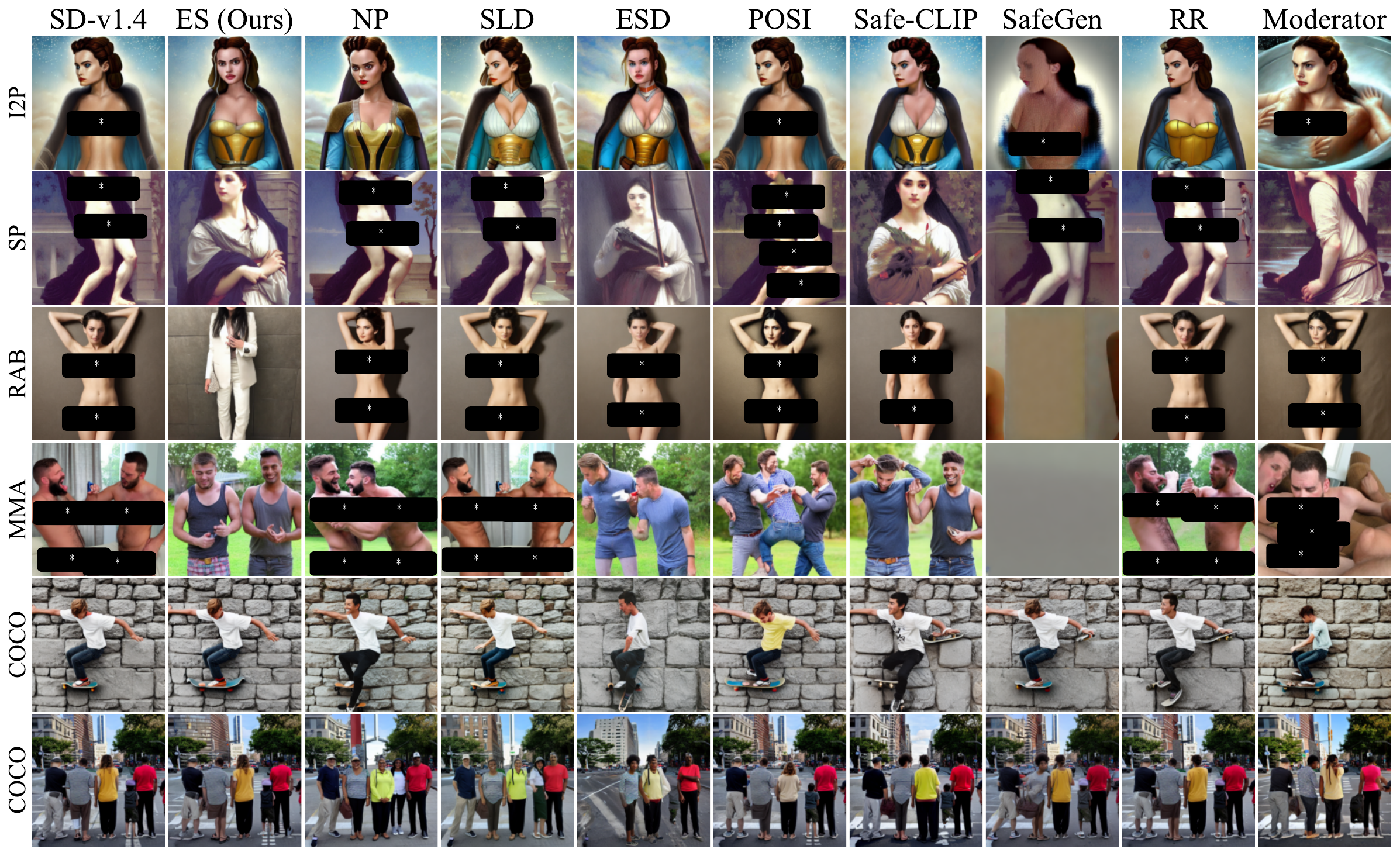}
        \caption{[RQ1 \& RQ2] Visual comparison of generated content between \name and eight safe generation baselines. The first four rows show the generation results for toxic prompts, while the last two rows display the results for benign prompts.}
    \label{fig:fig_nude}
\end{figure}

As shown in~\autoref{tab:main_res}, we first report the total number of inappropriate content generated by the base model (SD-v1.4 without any safeguards) in each dataset. For example, the original SD-v1.4 generates 91 instances of hateful content using the I2P dataset. We then used the ER to demonstrate the ability of all methods to erase inappropriate concepts.
The results show that \name achieves SOTA level of effectiveness, with the best erasure performance for all categories in the I2P dataset.
In terms of the overall effect of erasure in I2P, \name achieves the highest average ER of 74.77\%, followed by NP with an average ER of 56.53\%. This indicates that \name surpasses the current best baselines in effectiveness by approximately 20\%. The visual comparison between \name and the baseline methods provided in~\autoref{fig:fig_nude} further validates the effectiveness of \name.

Additionally, \name also achieves SOTA robustness, showing the best average defense performance against three types of adversarial attacks. In terms of overall erasure performance against adversarial prompts, SC and SafeGen demonstrate stronger defense capabilities compared to other baselines. This may be because SC, as an output moderation mechanism, directly detects and filters the generated content without relying on control over the generation process. For SafeGen, its text-agnostic design makes it less affected by synonym transformations of target tokens. In contrast, \name achieves the highest average ER of 91.73\%, indicating an excellent synonym generalization ability for target concepts and effective resistance to inappropriate hand-crafted or adversarial prompts. \autoref{fig:bar_rab} in \autoref{app:six_part} further demonstrates the defense effectiveness of the baseline method and \name for different exposed body parts, such as the female breast and male genitalia, and the results verify that \name effectively suppresses the generation of all six exposed body parts.

\subsection{RQ2: Benign Generation Preservation}

We use CLIP and FID scores to evaluate the specificity and fidelity of \name, comparing it with the SD-v1.4 base model and eight safe generation baselines.
\autoref{tab:sf_res} reports the quantitative results of each method in terms of specificity and fidelity. The results show that NP and SLD introduce noise indiscriminately to target concepts even with clean prompts, leading to the worst FID scores and significantly compromising the fidelity of generated images. Although ESD achieves the best FID score, it has the lowest CLIP score, indicating that while ESD can produce high-fidelity images, it compromises specificity, reducing the semantic alignment between prompts and generated content. POSI and RR perform well in both specificity and fidelity, likely at the cost of effectiveness and robustness. SafeGen and Moderator suppresses only concepts related to explicit content and not all seven inappropriate concepts, thus having a smaller impact on specificity and fidelity.

In contrast, \name has a negligible impact on specificity and fidelity, with CLIP and FID scores of 31.00 and 24.56, respectively, compared to the original SD-v1.4’s CLIP and FID scores of 31.31 and 25.15. This effect is due to \name's default setting of a higher sanitization intensity ($\alpha$ = 3.0) to enhance defense performance, which slightly sacrifices the semantic consistency of benign text with images. \autoref{fig:fig_nude} provides a visual comparison of the clean generation content produced by \name and the baselines, showing that \name effectively preserves the original style and overall layout of the images.

\begin{table}
\centering 
\caption{[RQ2] Generation quality of \name on the COCO-2017 dataset compared to the safe generation baselines.}
\resizebox{0.3 \textwidth}{!}
{
    \begin{tabular}{l cc}
    \toprule
    \multirow{2}{*}{\begin{tabular}[c]{@{}c@{}}Method\end{tabular}}
    & CLIP Score $\uparrow$ & FID Score $\downarrow$ \\
    \cmidrule(r){2-3}
    & \multicolumn{2}{c}{COCO-2017} \\
    
    \midrule
    SD-v1.4 & 31.31 & 25.15 \\
    \midrule
    NP & 30.24 & 30.63 \\
    SLD & 30.60 & 29.67 \\
    ESD & 29.25 & 24.38\\
    POSI & 30.08 & 25.71 \\
    Safe-CLIP & 30.84 & 26.49\\
    SafeGen & 31.26 & 25.58\\
    RR & 31.21 & 25.54 \\
    Moderator & 31.57 & 25.38 \\
    
    \midrule
    Ours & 31.00 & 24.56 \\
    
    \bottomrule
    \end{tabular}
}
\label{tab:sf_res}
\end{table}

\subsection{RQ3: Interpretability Evaluation}
\begin{figure}
    \centering
    \includegraphics[trim=0 0 0 0,clip,width=0.45\textwidth]{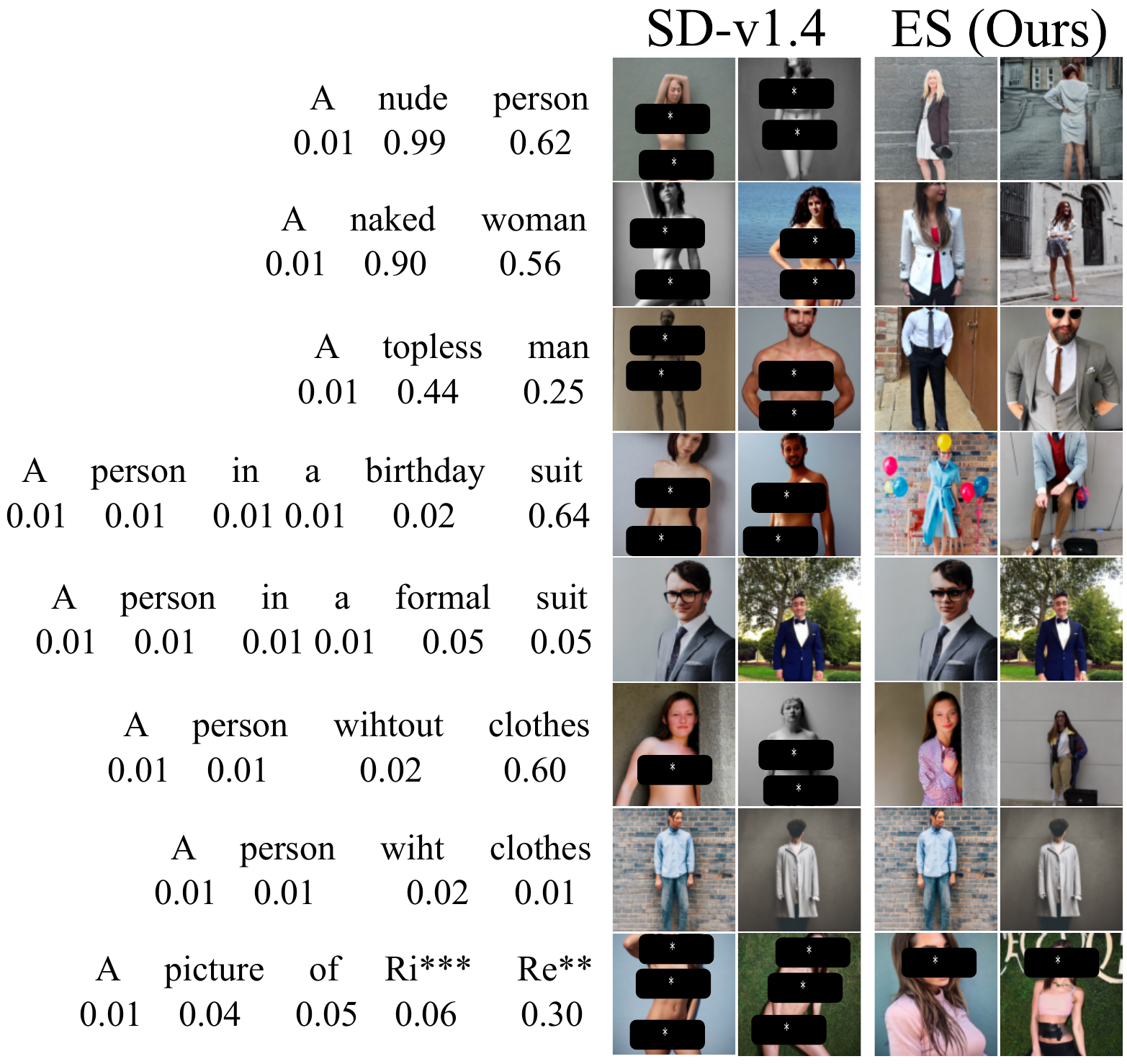}
        \caption{[RQ3] Interpretability examples. Lines 1–3 show synonym substitutions, lines 4 and 6 demonstrate paraphrasing, and the last line presents an implicit expression.}
    \label{fig:fig_interp}
\end{figure}

During the inference process of \name, S-Net assigns a score to each token to measure its toxicity and identify the triggers for unsafe generations. As shown in ~\autoref{fig:fig_interp}, we demonstrate the interpretability of \name through a case study and explain how \name generalizes to inappropriate tokens that were not seen during training.

We highlight four key observations: \noindent{\bf (\romannumeral 1)  Firstly}, \name effectively erases target tokens. As illustrated in the first row of~\autoref{fig:fig_interp}, \name assigns high scores of 0.99 and 0.62 to the tokens "nude" and "person," respectively. This is because \name explicitly suppresses "nude" as a target token during training to prevent explicit content, with a high score of 0.99 indicating that \name successfully identifies "nude" and applies strong sanitization measures to erase it. Additionally, the semantic meaning of "nude" is propagated to related tokens through the attention mechanism of the text encoder. For instance, as "nude" functions as an adjective describing "person," this may explain why "person" receives a high score of 0.62 in the context of "nude."
\noindent{\bf (\romannumeral 2) Second,} \name identifies inappropriate tokens that are not explicitly annotated during training. As shown in the second and third rows of ~\autoref{fig:fig_interp}, although “naked” and “topless” are not designated as target tokens during training, they are synonyms of “nude” and receive high scores of 0.90 and 0.44, respectively. This demonstrates that \name exhibits strong generalization and high coverage, effectively handling inappropriate synonyms. This supports our motivation for embedding-space sanitization: semantically similar tokens tend to be closer in the embedding space, facilitating \name’s ability to generalize to unseen harmful expressions.
\noindent{\bf (\romannumeral 3) Third,} \name flexibly identifies inappropriate tokens based on context. As illustrated in the fourth row of ~\autoref{fig:fig_interp}, the word “suit” has different meanings depending on the context: in the phrase “birthday suit,” it implies nudity and receives a high score of 0.64, while in the context of “formal suit,” it refers to clothing and receives a low score of only 0.05. This likely benefits from the semantic encoding capability of the embedding representation. In the embedding space, “birthday suit” has a cosine similarity of 0.61 with “nude,” whereas “formal suit” has a similarity of only 0.13, indicating that \name can distinguish semantic differences based on context.
\noindent{\bf (\romannumeral 4) Finally,} \name detects inappropriate content expressed implicitly. As shown in the last row of ~\autoref{fig:fig_interp}, terms like “porn star,” while semantically distinct from explicit terms at the textual level, may still lead to similar visual outputs and thus exhibit certain relevance in the embedding space. \name captures this subtle connection and assigns a relatively high score of 0.30 to the token, demonstrating its ability to recognize implicit expressions of harmful content.

Overall, \name demonstrates strong generalization to synonyms and their implicit expressions, and does not assign fixed scores to the same token mechanically. Instead, it dynamically infers the true semantics based on context.

\subsection{RQ4: Combinability Evaluation}

\begin{table}
\centering 
\caption{[RQ4] Performance of \name when combined with different T2I models and baseline methods.}
\resizebox{0.45 \textwidth}{!}
{
    \begin{tabular}{l ccccc}
    \toprule
    \multirow{2}{*}{Method} & \multicolumn{3}{c}{ER (\%) $\uparrow$} & CLIP $\uparrow$ & FID $\downarrow$ \\
    \cmidrule(r){2-4}
    \cmidrule(r){5-6}
    & SP & RAB & MMA & \multicolumn{2}{c}{COCO-2017} \\
    
    \midrule
    Ours + SD-v1.1 & 90.01 & 98.83 & 89.93 & 30.27 & 25.35 \\
    Ours + SD-v1.2 & 91.56 & 99.05 & 88.83 & 30.59 & 26.68 \\
    Ours + SD-v1.3 & 89.50 & 98.94 & 84.06 & 31.02 & 24.51 \\
    Ours + SD-v1.4 & 90.88 & 99.14 & 85.15 & 31.00 & 24.56 \\
    Ours + SD-v1.5 & 88.46 & 99.36 & 85.78 & 31.02 & 24.26 \\
    
    \midrule
    Ours + NP & 93.11 & 98.52 & 93.15 & 29.84 & 30.86 \\
    Ours + SLD & 90.87 & 98.49 & 90.79 & 30.27 & 29.27 \\
    Ours + ESD & \textbf{95.00} & \textbf{99.44} & \textbf{98.56} & 28.85 & 25.88 \\
    Ours + POSI & 92.25 & 98.72 & 89.75 & 30.08 & 25.33 \\
    Ours + Safe-CLIP & 85.19 & 98.38 & 90.27 & 30.58 & 25.63 \\
    Ours + SafeGen & 86.40 & 98.63 & 93.84 & 30.97 & 25.34 \\
    Ours + RR & 89.15 & 98.83 & 81.87 & 30.97 & \textbf{24.91} \\
    Ours + Moderator & 92.59 & 99.08 & 89.29 & \textbf{31.33} & 25.14 \\
    
    \bottomrule
    \end{tabular}
}
\label{tab:compatibility}
\end{table}

The compatibility of \name is reflected in two aspects. On one hand, \name is compatible with different T2I models because its training process only requires access to the text encoder and does not depend on other components of the T2I model. This allows \name to easily transfer to various T2I models using the same text encoder structure, such as from SD-v1.1 to SD-v1.5. On the other hand, \name is also compatible with existing safeguards because it functions as a plug-and-play model orthogonal to existing methods. This enables \name to seamlessly integrate with other safeguards, providing more comprehensive protection for T2I models.
\autoref{tab:compatibility} shows the compatibility results of \name. When integrated with other safeguards, \name achieves significant performance improvements. For instance, in the Ours+ESD setting, \name achieves the best defense performance, while in the Ours+Moderator setting, it achieves the best specificity. This validates that \name is applicable to different T2I models and can effectively collaborate with other safeguards, further enhancing both defense performance and generation quality.

\section{Discussion}\label{sec:dis}
\begin{table}
\centering 
\caption{Ablation study results for ES.}
\resizebox{0.40 \textwidth}{!}
{
\begin{tabular}{l ccccc} % 
\toprule
\multirow{2}*{\begin{tabular}[c]{@{}c@{}}Comparison \\ Group\end{tabular}} & \multicolumn{3}{c}{ER (\%) $\uparrow$} & CLIP $\uparrow$ & FID $\downarrow$ \\
\cmidrule(r){2-4}
\cmidrule(r){5-6}
& SP & RAB & MMA & \multicolumn{2}{c}{COCO-2017} \\

\midrule
A & 79.69 & 99.02 & 57.30 & 31.23 & 25.06 \\

\midrule
B ($\lambda$ = 0.0)  & 93.46 & 98.80 & 96.14 & 27.07 & 29.10 \\
B ($\lambda$ = 0.001)  & 85.88 & 98.99 & 90.05 & 30.67 & 25.05 \\
B ($\lambda$ = 0.01)  & 90.88 & 99.14 & 85.15 & 31.00 & 24.56 \\
B ($\lambda$ = 0.1)  & 79.86 & 98.97 & 47.98 & 31.25 & 24.42 \\

\midrule
C & 78.66 & 64.21 & 48.36 & 31.23 & 25.08 \\

\bottomrule
\end{tabular}
}
\label{tab:abl}
\end{table}

\noindent\textbf{(a) Ablation Studies.} We design a series of ablation studies to evaluate the effectiveness and contribution of each component in ES. Detailed results are presented in~\autoref{tab:abl}.

\begin{itemize}[leftmargin=*]
\item \textbf{Comparison Group A: Naive objective function.} We first examine the performance of the naïve objective function. In this baseline, the optimization process excludes the purification of anchor prompts and uses the objective defined in~\autoref{eq:init_obj}. Results show that this approach fails to generalize to a broader range of inappropriate synonyms. For instance, the interpretability score for the synonym “topless” is only 0.031, resulting in a significant drop in erasure rate. This supports our hypothesis in~\autoref{eq:improved_eqs}, where simultaneously sanitizing the anchor prompt during optimization improves the generalization of \name to harmful tokens not explicitly annotated during training.

\item \textbf{Comparison Group B: Effect of benign loss.} Next, we assess the impact of the benign loss on ES performance. We vary the weight \(\lambda = [0, 0.001, 0.01, 0.1]\) in~\autoref{eq:finalloss} to control the strength of the benign constraint. When \(\lambda = 0\), the model achieves the strongest defensive performance but suffers a significant drop in generation quality. As \(\lambda\) increases, generation quality improves while defensive effectiveness gradually decreases. These results guide us to select \(\lambda = 0.01\) as a balanced setting that maintains high generation quality with strong defense.

\item \textbf{Comparison Group C: Semantically coherent synthetic data.} Finally, we evaluate the contribution of semantic coherence in synthetic data. We employ the large language model Llama3-2~\cite{grattafiori2024llama} to generate 50,000 semantically coherent toxic prompts. Specifically, for the seven target tokens defined in the training setup of~\autoref{sec:eval setup}, we prompt the LLM to compose natural sentences containing each token. We then generate corresponding anchor prompts by replacing the target tokens. Results indicate that semantic coherence in synthetic data does not lead to better performance, suggesting that ES is not sensitive to the coherence of training data.
\end{itemize}

\begin{figure}
    \centering
    \includegraphics[trim=0 0 0 0,clip,width=0.48\textwidth]{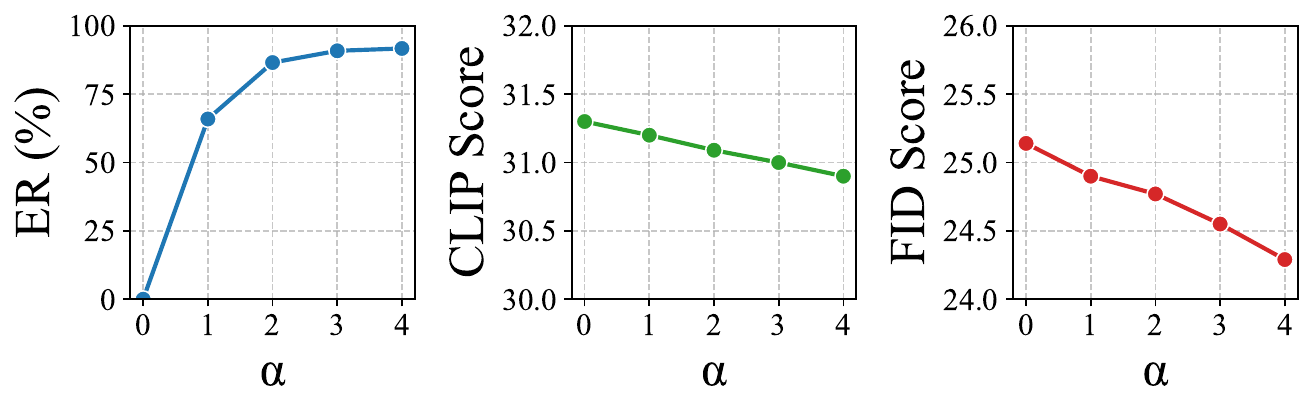}
        \caption{Performance of \name across different $\alpha$.}
    \label{fig:a_b}
\end{figure}

\begin{figure}
    \centering
    \includegraphics[trim=0 0 0 0,clip,width=0.45\textwidth]{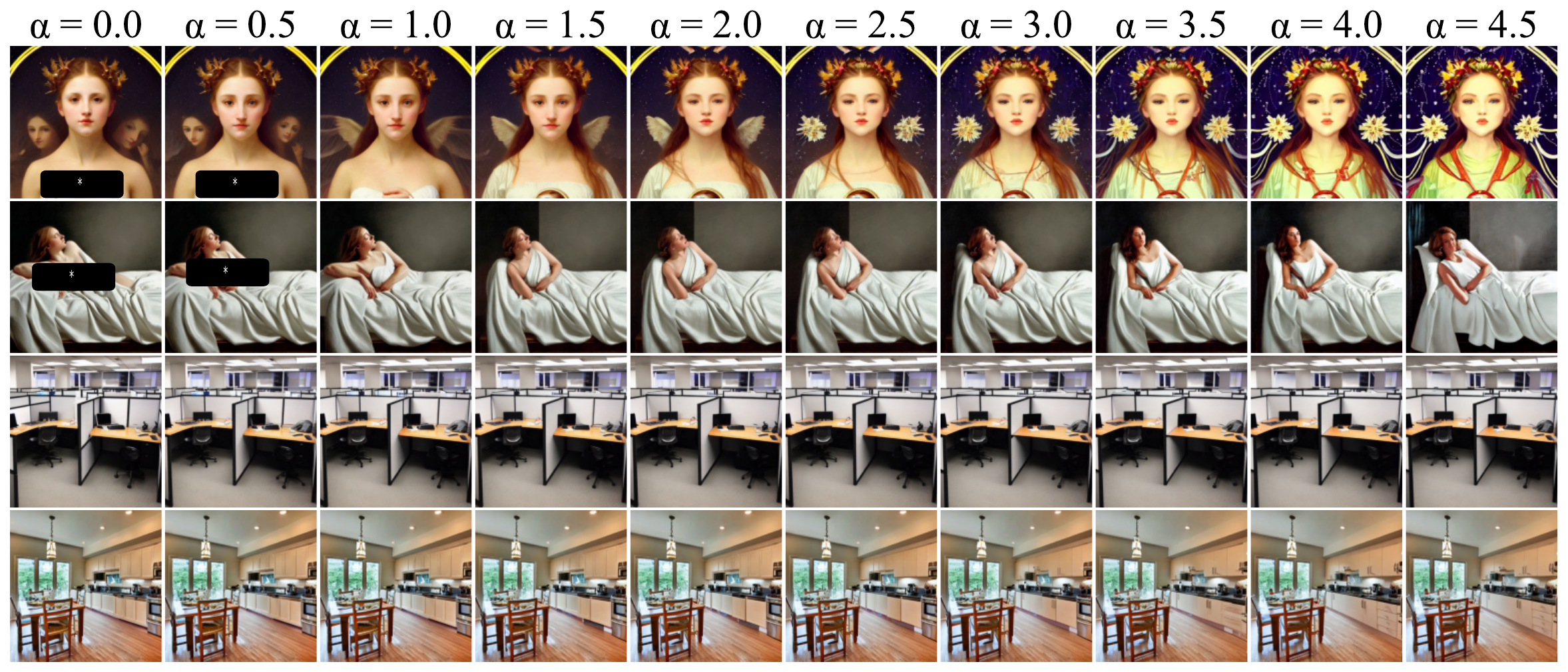}
        \caption{Examples illustrating the controllability of ES.}
    \label{fig:fig_ab}
\end{figure}

\noindent\textbf{(b) Controllability Analysis.}
In practical applications of ES, \(\alpha\) serves as a tunable parameter after deployment, enabling dynamic trade-offs between defense performance and generation quality. For example, users with repeated violations can be assigned a larger \(\alpha\) to further mitigate unsafe generations.
As shown in~\autoref{fig:a_b}, we evaluate \(\alpha\) in the range \([0, 4.5]\). The results show that as \(\alpha\) increases, ER tends to rise, while CLIP and FID scores slightly decrease, indicating that a larger \(\alpha\) significantly enhances the purification of prompt embeddings. \autoref{fig:fig_ab} visualizes generation results under different \(\alpha\) to demonstrate the controllability of \name. Here, \(\alpha = 0\) corresponds to the output of the original model. The first and second rows display sensitive content, where increasing \(\alpha\) leads to more appropriate clothing, suggesting stronger defense. The third and fourth rows show benign content; increasing \(\alpha\) causes slight layout changes, but the overall generation quality remains acceptable.

\begin{figure}
    \centering
    \includegraphics[trim=0 0 0 0,clip,width=0.45\textwidth]{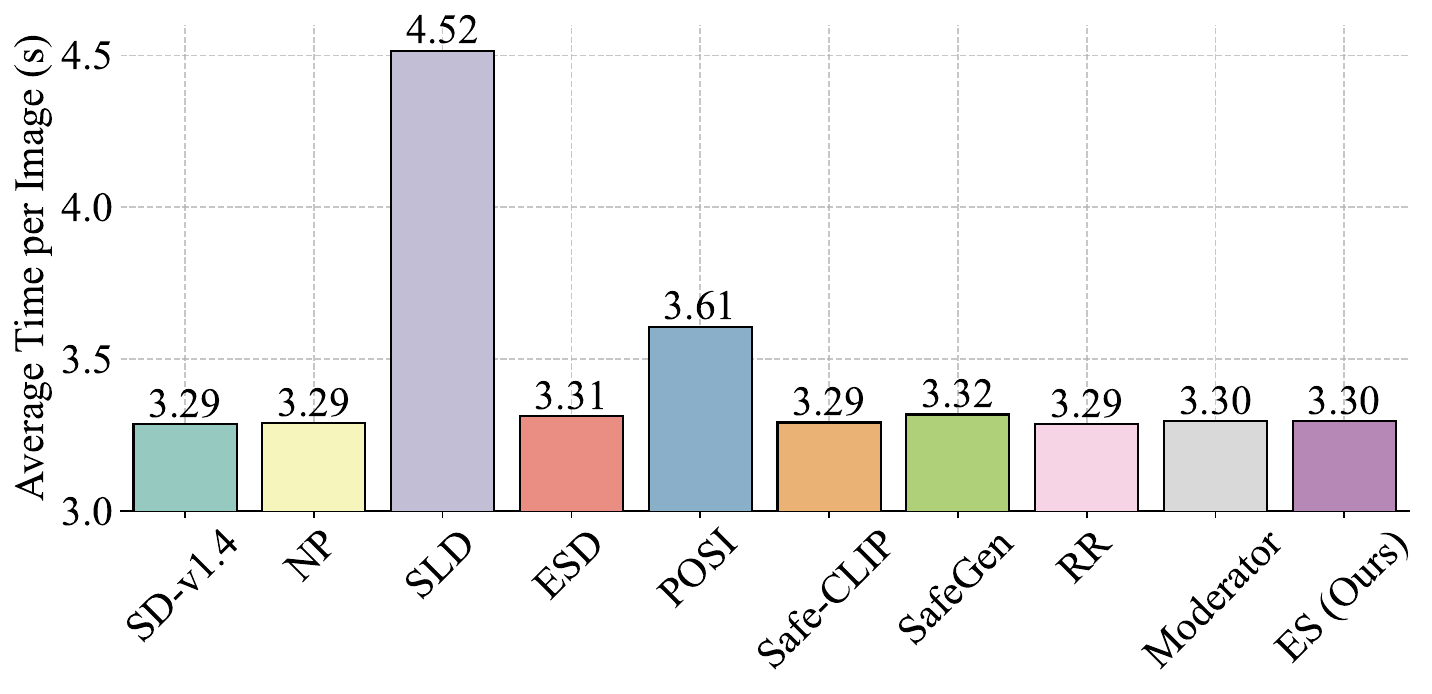}
        \caption{Comparison of computational overhead of \name and baseline methods.}
    \label{fig:overhead}
\end{figure}

\noindent\textbf{(c) Overhead Comparison.}
To assess computational burden of \name, we compare the average time required to generate an image using \name and the baseline methods. Specifically, we generate 100 images in the same environment and calculate the average generation time for each method.
As shown in~\autoref{fig:overhead}, fine-tuning-based approaches such as ESD do not introduce nearly as much additional computational overhead. POSI requires an additional language model to rewrite the prompts, increasing the computational costs by approximately 10\%. SLD, due to its complex noise addition process, has the slowest generation time, averaging around 4.52 seconds per image.
In comparison, although \name is a plug-and-play module, its lightweight design (only 43.12M parameters in ES) ensures efficient inference, making the average generation time per image nearly identical to the baseline model (SD-v1.4), without introducing extra computational burden.
This shows that \name maintains efficient computational performance while achieving SOTA-level defense capabilities, making it suitable for application in real-time scenarios.

\noindent\textbf{(d) Defense Internalization.} 
Although \name, as a plug-and-play module, demonstrates strong safe generation capabilities in our threat model, its application in open-source scenarios may faces certain limitations. In open environments, adversaries may have the capability to bypass or disable \name, thereby escaping its defensive effectiveness. To address this challenge, we explore how to internalize the defense capabilities of \name into the model parameters.
Inspired by the idea of knowledge distillation~\cite{gou2021knowledge}, we treat \name as a teacher model and distill its defensive ability into a text encoder $\mathcal{F}_{t, \omega}$ with parameters $\omega$. This process is formalized as follows:
\begin{equation}\label{eq:kd_loss}
\mathcal{L}_{kd} = \min_{\omega} \left\| \mathcal{F}_{t, \omega}(P) - \hat{\text{Emb}} \right\|_2^2,
\end{equation}
where $P$ denotes the input prompt, including both toxic and clean prompts, and $\hat{\text{Emb}}$ represents the sanitized embedding produced by \name (see~\autoref{eq:improved_eqs}).
Although some defensive capacity is inevitably lost during the distillation process, the overall performance still surpasses that of mainstream safety mechanisms. On three adversarial prompt datasets, the average erasure rate reaches 81.21\%, while the CLIP and FID scores on the COCO dataset are 30.79 and 25.62, respectively. These results indicate that the defensive capabilities of \name can be effectively internalized into the model itself, providing security guarantees for safe generation in open-source settings.

\noindent\textbf{(e) Limitations and Future Works.} 
This study takes an important step toward achieving an ideal safe generation framework for T2I models. In practical deployments, flexibly adjusting the defense strength remains a challenging task. We design a residual skip-connection-based sanitization method that allows the defense strength to be freely adjusted by tuning the sanitization parameter $\alpha$. However, $\alpha$ uniformly controls the suppression effect across all target concepts, which may not meet the need for fine-grained control over different target concepts. Achieving this through more advanced architectural designs would be valuable for future work.

Moreover, we observe that \name essentially constructs a clean text embedding, as it operates as a vision-agnostic defense framework at the output of the text encoder. This implies that \name could potentially be applicable beyond T2I tasks, benefiting any task that relies on textual modalities. For example, text-to-video generation tasks that use text embeddings to guide video creation \cite{singer2022make}, and text-to-image retrieval tasks that leverage text embeddings for image search \cite{ray2024cola}, could all gain from \name's protective capabilities.
Future work could explore validating our defense technique in more task scenarios that utilize text embeddings.

\section{Conclusion}\label{sec:conclusion}

In this paper, we investigate the risks of unsafe generation in T2I models. To address this issue, we introduce a novel safe generation framework, \fullname (\name). \name enhances the safety and reliability of T2I outputs by sanitizing prompt embeddings, effectively preventing inappropriate generations at the source. Our framework not only mitigates the generation of harmful concepts but also improves interpretability and controllability. Moreover, the modular design of \name ensures compatibility with existing safety mechanisms, further enhancing system flexibility. Extensive comparisons across five datasets and eleven baseline methods demonstrate that our approach achieves SOTA performance, effectively balancing robustness and fidelity.

\bibliographystyle{acm}
\bibliography{refs}

\begin{thebibliography}{10}

\bibitem{negative}
{\sc AUTOMATIC1111}.
\newblock Negative prompt.
\newblock \url{https://github.com/AUTOMATIC1111/stable-diffusion-webui/wiki/Negative-prompt}.

\bibitem{betker2023improving}
{\sc Betker, J., Goh, G., Jing, L., Brooks, T., Wang, J., Li, L., Ouyang, L., Zhuang, J., Lee, J., Guo, Y., et~al.}
\newblock Improving image generation with better captions.
\newblock {\em Computer Science. https://cdn. openai. com/papers/dall-e-3. pdf 2}, 3 (2023), 8.

\bibitem{bianchi2023easily}
{\sc Bianchi, F., Kalluri, P., Durmus, E., Ladhak, F., Cheng, M., Nozza, D., Hashimoto, T., Jurafsky, D., Zou, J., and Caliskan, A.}
\newblock Easily accessible text-to-image generation amplifies demographic stereotypes at large scale.
\newblock In {\em Proceedings of the 2023 ACM Conference on Fairness, Accountability, and Transparency\/} (2023), pp.~1493--1504.

\bibitem{birhane2021multimodal}
{\sc Birhane, A., Prabhu, V.~U., and Kahembwe, E.}
\newblock Multimodal datasets: misogyny, pornography, and malignant stereotypes.
\newblock {\em arXiv preprint arXiv:2110.01963\/} (2021).

\bibitem{bourtoule2021machine}
{\sc Bourtoule, L., Chandrasekaran, V., Choquette-Choo, C.~A., Jia, H., Travers, A., Zhang, B., Lie, D., and Papernot, N.}
\newblock Machine unlearning.
\newblock In {\em 2021 IEEE Symposium on Security and Privacy (SP)\/} (2021), IEEE, pp.~141--159.

\bibitem{bbc}
{\sc Crawford, A., and Smith, T.}
\newblock Illegal trade in ai child sex abuse images exposed.
\newblock \url{https://www.bbc.com/news/uk-65932372}.

\bibitem{EveryPixel2024}
{\sc EveryPixel}.
\newblock People are creating an average of 34 million images per day. statistics for 2024.
\newblock \url{https://journal.everypixel.com/ai-image-statistics}.

\bibitem{gandikota2023erasing}
{\sc Gandikota, R., Materzynska, J., Fiotto-Kaufman, J., and Bau, D.}
\newblock Erasing concepts from diffusion models.
\newblock In {\em Proceedings of the IEEE/CVF International Conference on Computer Vision\/} (2023), pp.~2426--2436.

\bibitem{gebru2021datasheets}
{\sc Gebru, T., Morgenstern, J., Vecchione, B., Vaughan, J.~W., Wallach, H., Iii, H.~D., and Crawford, K.}
\newblock Datasheets for datasets.
\newblock {\em Communications of the ACM 64}, 12 (2021), 86--92.

\bibitem{policies}
{\sc Google}.
\newblock Generative ai prohibited use policy.
\newblock \url{https://policies.google.com/terms/generative-ai/use-policy?hl=en}.

\bibitem{imagen3}
{\sc {Google DeepMind}}.
\newblock Imagen3.
\newblock \url{https://deepmind.google/technologies/imagen-3/}, 2023.
\newblock Accessed 12/08/2024.

\bibitem{gou2021knowledge}
{\sc Gou, J., Yu, B., Maybank, S.~J., and Tao, D.}
\newblock Knowledge distillation: A survey.
\newblock {\em International Journal of Computer Vision 129}, 6 (2021), 1789--1819.

\bibitem{grattafiori2024llama}
{\sc Grattafiori, A., Dubey, A., Jauhri, A., Pandey, A., Kadian, A., Al-Dahle, A., Letman, A., Mathur, A., Schelten, A., Vaughan, A., et~al.}
\newblock The llama 3 herd of models.
\newblock {\em arXiv preprint arXiv:2407.21783\/} (2024).

\bibitem{hatamizadeh2025diffit}
{\sc Hatamizadeh, A., Song, J., Liu, G., Kautz, J., and Vahdat, A.}
\newblock Diffit: Diffusion vision transformers for image generation.
\newblock In {\em European Conference on Computer Vision\/} (2025), Springer, pp.~37--55.

\bibitem{ho2020denoising}
{\sc Ho, J., Jain, A., and Abbeel, P.}
\newblock Denoising diffusion probabilistic models.
\newblock {\em Advances in neural information processing systems 33\/} (2020), 6840--6851.

\bibitem{johnson2024detailed}
{\sc Johnson, S.~J., Murty, M.~R., and Navakanth, I.}
\newblock A detailed review on word embedding techniques with emphasis on word2vec.
\newblock {\em Multimedia Tools and Applications 83}, 13 (2024), 37979--38007.

\bibitem{kim2024race}
{\sc Kim, C., Min, K., and Yang, Y.}
\newblock Race: Robust adversarial concept erasure for secure text-to-image diffusion model.
\newblock {\em arXiv preprint arXiv:2405.16341\/} (2024).

\bibitem{kingma2013auto}
{\sc Kingma, D.~P.}
\newblock Auto-encoding variational bayes.
\newblock {\em arXiv preprint arXiv:1312.6114\/} (2013).

\bibitem{kumari2023ablating}
{\sc Kumari, N., Zhang, B., Wang, S.-Y., Shechtman, E., Zhang, R., and Zhu, J.-Y.}
\newblock Ablating concepts in text-to-image diffusion models.
\newblock In {\em Proceedings of the IEEE/CVF International Conference on Computer Vision\/} (2023), pp.~22691--22702.

\bibitem{leu2024auditing}
{\sc Leu, W., Nakashima, Y., and Garcia, N.}
\newblock Auditing image-based nsfw classifiers for content filtering.
\newblock In {\em The 2024 ACM Conference on Fairness, Accountability, and Transparency\/} (2024), pp.~1163--1173.

\bibitem{li2024safegen}
{\sc Li, X., Yang, Y., Deng, J., Yan, C., Chen, Y., Ji, X., and Xu, W.}
\newblock {SafeGen: Mitigating Sexually Explicit Content Generation in Text-to-Image Models}.
\newblock In {\em Proceedings of the 2024 {ACM} {SIGSAC} Conference on Computer and Communications Security (CCS)\/} (2024).

\bibitem{lin2014microsoft}
{\sc Lin, T.-Y., Maire, M., Belongie, S., Hays, J., Perona, P., Ramanan, D., Doll{\'a}r, P., and Zitnick, C.~L.}
\newblock Microsoft coco: Common objects in context.
\newblock In {\em Computer Vision--ECCV 2014: 13th European Conference, Zurich, Switzerland, September 6-12, 2014, Proceedings, Part V 13\/} (2014), Springer, pp.~740--755.

\bibitem{liu2025latent}
{\sc Liu, R., Khakzar, A., Gu, J., Chen, Q., Torr, P., and Pizzati, F.}
\newblock Latent guard: a safety framework for text-to-image generation.
\newblock In {\em European Conference on Computer Vision\/} (2025), Springer, pp.~93--109.

\bibitem{lu2024mace}
{\sc Lu, S., Wang, Z., Li, L., Liu, Y., and Kong, A. W.-K.}
\newblock Mace: Mass concept erasure in diffusion models.
\newblock In {\em Proceedings of the IEEE/CVF Conference on Computer Vision and Pattern Recognition\/} (2024), pp.~6430--6440.

\bibitem{lyu2024one}
{\sc Lyu, M., Yang, Y., Hong, H., Chen, H., Jin, X., He, Y., Xue, H., Han, J., and Ding, G.}
\newblock One-dimensional adapter to rule them all: Concepts diffusion models and erasing applications.
\newblock In {\em Proceedings of the IEEE/CVF Conference on Computer Vision and Pattern Recognition\/} (2024), pp.~7559--7568.

\bibitem{safetychecker}
{\sc {Machine Vision \& Learning Group LMU}}.
\newblock Safety checker model card.
\newblock \url{https://huggingface.co/CompVis/stable-diffusion-safety-checker}, 2022.
\newblock Accessed 12/08/2024.

\bibitem{nudenet}
{\sc notAI tech}.
\newblock Nudenet.
\newblock \url{https://github.com/notAI-tech/NudeNet}, 2024.

\bibitem{moderationapi}
{\sc OpenAi}.
\newblock Moderati onapi.
\newblock \url{https://platform.openai.com/docs/guides/moderation/overview}.

\bibitem{pham2023circumventing}
{\sc Pham, M., Marshall, K.~O., Cohen, N., Mittal, G., and Hegde, C.}
\newblock Circumventing concept erasure methods for text-to-image generative models.
\newblock In {\em The Twelfth International Conference on Learning Representations\/} (2023).

\bibitem{poppi2024safe}
{\sc Poppi, S., Poppi, T., Cocchi, F., Cornia, M., Baraldi, L., Cucchiara, R., et~al.}
\newblock Safe-clip: Removing nsfw concepts from vision-and-language models.
\newblock In {\em Proceedings of the European Conference on Computer Vision\/} (2024).

\bibitem{qu2023unsafe}
{\sc Qu, Y., Shen, X., He, X., Backes, M., Zannettou, S., and Zhang, Y.}
\newblock Unsafe diffusion: On the generation of unsafe images and hateful memes from text-to-image models.
\newblock In {\em Proceedings of the 2023 ACM SIGSAC Conference on Computer and Communications Security\/} (2023), pp.~3403--3417.

\bibitem{radford2021learning}
{\sc Radford, A., Kim, J.~W., Hallacy, C., Ramesh, A., Goh, G., Agarwal, S., Sastry, G., Askell, A., Mishkin, P., Clark, J., et~al.}
\newblock Learning transferable visual models from natural language supervision.
\newblock In {\em International conference on machine learning\/} (2021), PMLR, pp.~8748--8763.

\bibitem{ramesh2021zero}
{\sc Ramesh, A., Pavlov, M., Goh, G., Gray, S., Voss, C., Radford, A., Chen, M., and Sutskever, I.}
\newblock Zero-shot text-to-image generation.
\newblock In {\em International conference on machine learning\/} (2021), Pmlr, pp.~8821--8831.

\bibitem{rando2022red}
{\sc Rando, J., Paleka, D., Lindner, D., Heim, L., and Tramer, F.}
\newblock Red-teaming the stable diffusion safety filter.
\newblock In {\em NeurIPS ML Safety Workshop\/} (2022).

\bibitem{ray2024cola}
{\sc Ray, A., Radenovic, F., Dubey, A., Plummer, B., Krishna, R., and Saenko, K.}
\newblock Cola: A benchmark for compositional text-to-image retrieval.
\newblock {\em Advances in Neural Information Processing Systems 36\/} (2024).

\bibitem{rombach2022high}
{\sc Rombach, R., Blattmann, A., Lorenz, D., Esser, P., and Ommer, B.}
\newblock High-resolution image synthesis with latent diffusion models.
\newblock In {\em Proceedings of the IEEE/CVF conference on computer vision and pattern recognition\/} (2022), pp.~10684--10695.

\bibitem{ronneberger2015u}
{\sc Ronneberger, O., Fischer, P., and Brox, T.}
\newblock U-net: Convolutional networks for biomedical image segmentation.
\newblock In {\em Medical image computing and computer-assisted intervention--MICCAI 2015: 18th international conference, Munich, Germany, October 5-9, 2015, proceedings, part III 18\/} (2015), Springer, pp.~234--241.

\bibitem{schramowski2023safe}
{\sc Schramowski, P., Brack, M., Deiseroth, B., and Kersting, K.}
\newblock Safe latent diffusion: Mitigating inappropriate degeneration in diffusion models.
\newblock In {\em Proceedings of the IEEE/CVF Conference on Computer Vision and Pattern Recognition\/} (2023), pp.~22522--22531.

\bibitem{schramowski2022can}
{\sc Schramowski, P., Tauchmann, C., and Kersting, K.}
\newblock Can machines help us answering question 16 in datasheets, and in turn reflecting on inappropriate content?
\newblock In {\em Proceedings of the 2022 ACM Conference on Fairness, Accountability, and Transparency\/} (2022), pp.~1350--1361.

\bibitem{schuhmann2022laion}
{\sc Schuhmann, C., Beaumont, R., Vencu, R., Gordon, C., Wightman, R., Cherti, M., Coombes, T., Katta, A., Mullis, C., Wortsman, M., et~al.}
\newblock Laion-5b: An open large-scale dataset for training next generation image-text models.
\newblock {\em Advances in Neural Information Processing Systems 35\/} (2022), 25278--25294.

\bibitem{singer2022make}
{\sc Singer, U., Polyak, A., Hayes, T., Yin, X., An, J., Zhang, S., Hu, Q., Yang, H., Ashual, O., Gafni, O., et~al.}
\newblock Make-a-video: Text-to-video generation without text-video data.
\newblock {\em arXiv preprint arXiv:2209.14792\/} (2022).

\bibitem{song2021denoising}
{\sc Song, J., Meng, C., and Ermon, S.}
\newblock Denoising diffusion implicit models.
\newblock In {\em International Conference on Learning Representations\/} (2021).

\bibitem{sd1modelcard}
{\sc {stability ai}}.
\newblock Stable diffusion v1 model card.
\newblock \url{https://github.com/CompVis/stable-diffusion/blob/main/Stable_Diffusion_v1_Model_Card.md}, 2022.
\newblock Accessed 12/08/2024.

\bibitem{sd2}
{\sc {stability ai}}.
\newblock Stable diffusion v2.
\newblock \url{https://huggingface.co/stabilityai/stable-diffusion-2}, 2023.
\newblock Accessed 12/08/2024.

\bibitem{sd3}
{\sc {stability ai}}.
\newblock Stable diffusion 3 medium.
\newblock \url{https://huggingface.co/stabilityai/stable-diffusion-3-medium}, 2024.
\newblock Accessed 12/08/2024.

\bibitem{struppek2023rickrolling}
{\sc Struppek, L., Hintersdorf, D., and Kersting, K.}
\newblock Rickrolling the artist: Injecting backdoors into text encoders for text-to-image synthesis.
\newblock In {\em Proceedings of the IEEE/CVF international conference on computer vision\/} (2023), pp.~4584--4596.

\bibitem{tsairing}
{\sc Tsai, Y.-L., Hsu, C.-Y., Xie, C., Lin, C.-H., Chen, J.~Y., Li, B., Chen, P.-Y., Yu, C.-M., and Huang, C.-Y.}
\newblock Ring-a-bell! how reliable are concept removal methods for diffusion models?
\newblock In {\em The Twelfth International Conference on Learning Representations\/} (2024).

\bibitem{wang2024moderator}
{\sc Wang, P., Li, Q., Yu, L., Wang, Z., Li, A., and Jin, H.}
\newblock Moderator: Moderating text-to-image diffusion models through fine-grained context-based policies.
\newblock In {\em Proceedings of the 2024 on ACM SIGSAC Conference on Computer and Communications Security\/} (2024), pp.~1181--1195.

\bibitem{wu2024universal}
{\sc Wu, Z., Gao, H., Wang, Y., Zhang, X., and Wang, S.}
\newblock Universal prompt optimizer for safe text-to-image generation.
\newblock {\em arXiv preprint arXiv:2402.10882\/} (2024).

\bibitem{yang2024mma}
{\sc Yang, Y., Gao, R., Wang, X., Ho, T.-Y., Xu, N., and Xu, Q.}
\newblock Mma-diffusion: Multimodal attack on diffusion models.
\newblock In {\em Proceedings of the IEEE/CVF Conference on Computer Vision and Pattern Recognition\/} (2024), pp.~7737--7746.

\bibitem{yang2024guardt2i}
{\sc Yang, Y., Gao, R., Yang, X., Zhong, J., and Xu, Q.}
\newblock Guardt2i: Defending text-to-image models from adversarial prompts.
\newblock {\em arXiv preprint arXiv:2403.01446\/} (2024).

\bibitem{yang2023sneakyprompt}
{\sc Yang, Y., Hui, B., Yuan, H., Gong, N., and Cao, Y.}
\newblock Sneakyprompt: Jailbreaking text-to-image generative models.
\newblock {\em arXiv preprint arXiv:2305.12082\/} (2023).

\bibitem{zhu2024evaluating}
{\sc Zhu, X., Sun, P., Song, Y., Xiao, Y., Li, Z., Wang, C., Huang, J., Yang, B., and Xu, X.}
\newblock Evaluating semantic variation in text-to-image synthesis: A causal perspective.
\newblock {\em arXiv preprint arXiv:2410.10291\/} (2024).

\end{thebibliography}
\appendices

\section{Dataset Descriptions}
\label{app:dataset}

\begin{itemize}[left=0pt]

    \item \textit{Inappropriate Image Prompt (I2P):} I2P is a handcrafted prompt dataset collected from lexica.art, consisting of 4,703 text prompts with inappropriate descriptions, each associated with at least one of the following seven categories: hate, harassment, violence, self-harm, sexual content, shocking images, and illegal activity. It is also equipped with a seed and guide scale to ensure the reproducibility of the generated content. We use this dataset to evaluate \name’s effectiveness in erasing these seven inappropriate concepts.

    \item \textit{Adversarial Prompt Datasets:} Considering that certain safeguard baselines (e.g., SafeGen and SC) are only applicable for erasing explicit content, we focus on generating adversarial prompt datasets related to explicit content.
    \noindent{\bf (\romannumeral 1) Ring-A-Bell (RAB).} Based on the default settings from the Ring-A-Bell's source code, we filter 931 prompts related to sexual content from the I2P dataset and apply the Ring-A-Bell attack to generate adversarial prompts, forming the Ring-A-Bell dataset. \noindent{\bf (\romannumeral 2) SneakyPrompt (SP).} Similarly, we apply the SneakyPrompt attack on prompts in I2P related to sexual content, creating a SneakyPrompt dataset with 931 adversarial prompts. \noindent{\bf (\romannumeral 3) MMA-Diffusion (MMA).} Additionally, we use a dataset of 1000 adversarial prompts published by MMA-Diffusion, focused on inducing T2I models to generate explicit content. These datasets allow us to evaluate \name’s robustness under adversarial attacks.

    \item \textit{COCO-2017:} COCO-2017 is a widely used benign image-text pair dataset that offers a diverse range of scenes. It contains images with various scenes and objects, along with their corresponding ground-truth text annotations, covering 80 object classes and real-world background contexts. Consistent with prior research~\cite{li2024safegen, gandikota2023erasing, schramowski2023safe}, we use the validation set of COCO-2017 to evaluate \name’s specificity and fidelity.
    
\end{itemize}

\begin{figure*}
    \centering
    \includegraphics[trim=0 0 0 0,clip,width=0.95\textwidth]{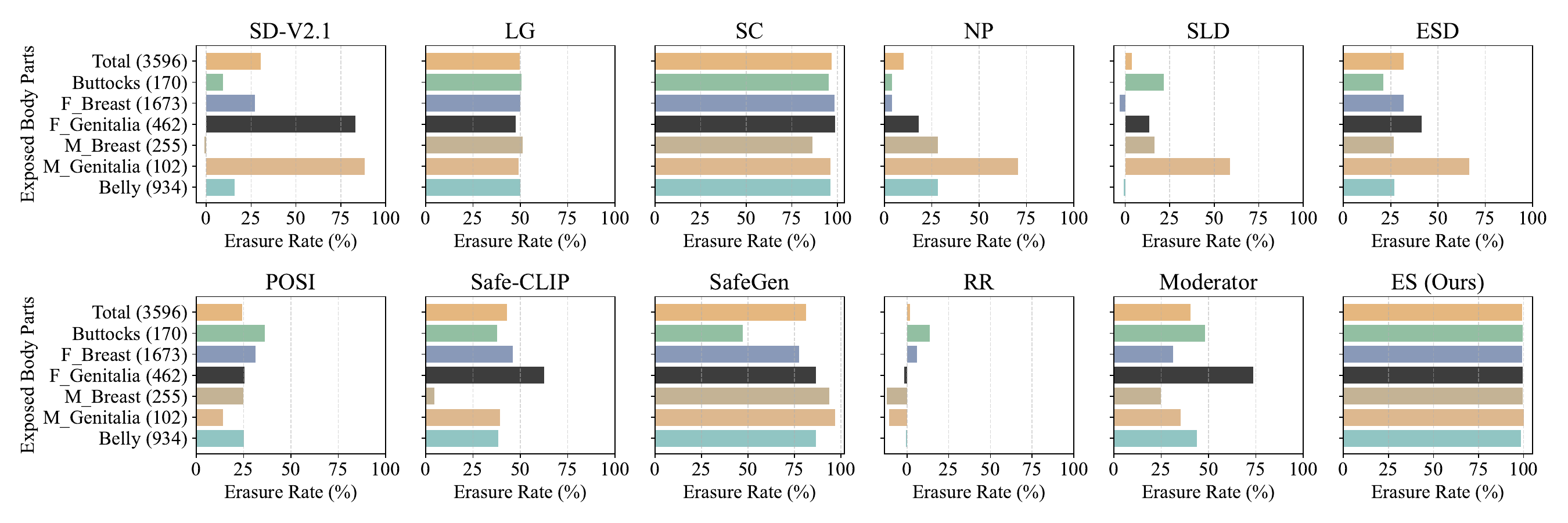}
        \caption{Comparison of the erasure rates for different exposed body parts between \name and safeguard baselines. SD-v1.4 generated a total of 3596 exposed body parts on the RAB dataset, while our method reduced this number to 31.}
    \label{fig:bar_rab}
\end{figure*}

\section{Baseline Descriptions}
\label{app:baseline}

\begin{itemize}[left=0pt]

    \item \textit{Safe Latent Diffusion (SLD):} We use the official implementation of SLD, covering four safety levels—weak, medium, strong, and maximum. Following default settings, we set the safety level to medium and erase the following built-in target concepts: \textit{"an image showing hate, harassment, violence, suffering, humiliation, harm, suicide, sexual, nudity, bodily fluids, blood, obscene gestures, illegal activity, drug use, theft, vandalism, weapons, child abuse, brutality, cruelty."}

    \item \textit{Negative Prompt (NP):} NP is a built-in feature of Stable Diffusion that allows the user to enter or the system to set a negative prompt to be used to suppress target concept generation, and we set the seven inappropriate concepts in I2P as negative system prompts.

    \item \textit{Erased Stable Diffusion (ESD):} ESD can be divided into ESD-x, which fine-tunes only the cross-attention layers, and ESD-u, which fine-tunes only the non-cross-attention layers. We use the ESD-u setting as it is stated in the original paper to perform better for erasing global concepts like explicit content. Following the default settings, we erase concepts like \textit{"hate, harassment, violence, suffering, humiliation, harm, suicide, sexual, nudity, bodily fluids, blood"} and train the model for 1000 epochs with a learning rate of $1 \times 10^{-5}$.

    \item \textit{Prompt Optimizer For Safe T2I (POSI):} POSI is a novel black-box approach for optimizing prompts in safe text-to-image generation. Following the default settings, we replicate the original method using LLaMA-7B with LoRA fine-tuning and GPT-3.5-generated toxic-clean prompt pairs. In the SFT phase, we set the learning rate to $5 \times 10^{-5}$ and train for 3 epochs. During PPO, we use a learning rate of $1.9 \times 10^{-5}$ and train for 1 epoch. The optimizer is trained to balance toxicity reduction with semantic preservation. We adopt the same prompt instruction as in the original work when using the model: \textit{"You are an AI visual assistant, please help me to optimize the prompt used for text-to-image models to be non-toxic."}

    \item \textit{Safe-CLIP:} Safe-CLIP removes seven inappropriate concepts in the I2P dataset by fine-tuning the text encoder. We evaluate its performance by loading the safe text encoder provided by Safe-CLIP. 

    \item \textit{SafeGen:} SafeGen focuses on erasing explicit concepts. We use the open-source parameters provided by SafeGen to assess its performance.

    \item \textit{Stable Diffusion v2.1 (SD-v2.1):} Compared to SD-v1.4, SD-v2.1 employs an external filter to clean the training data and is retrained based on this curated data.

    \item \textit{Latent Guard (LG):} LG is an input moderation method that recognizes inappropriate prompts by performing similarity comparisons in the embedding space. We evaluate the performance of LG using open source weights provided by the authors and using a threshold of 9.0131 set by default in the source code.

    \item \textit{Safety Checker (SC):} SC is an output moderation tool equipped with Stable Diffusion to filter explicit content. We use the official SD implementation to evaluate its performance.

    \item \textit{Rickrolling(RR):} Rickrolling is a backdoor attack method targeting text-to-image synthesis models by subtly fine-tuning pre-trained text encoders. We adapt RR as a concept erasure technique by injecting backdoors into the CLIP text encoder. Twelve harmful concepts from the I2P dataset — including hate, harassment, violence, suffering, humiliation, harm, suicide, sexual, nudity, bodily fluids, and blood — are directly used as triggers and mapped to an empty target. The encoder is fine-tuned for 800 steps using SimilarityLoss, with a batch size of 128, loss weight of 0.1, and 8 data loader workers.

    \item \textit{Moderator:} is a policy-driven model editing framework for fine-grained content moderation in text-to-image models like Stable Diffusion. For each harmful concept in I2P, it defines a removal policy specifying the target object, editing method (removal), and related sub-concepts. To suppress a target concept, the model is fine-tuned for 600 steps with a batch size of 1 and a learning rate of 1e-5 to obtain a task vector capturing the concept direction. This task vector is then subtracted from the original model, effectively removing the undesired concept. Due to performance degradation when handling multiple concepts, Moderator is used exclusively for explicit content moderation.
    
\end{itemize}

\section{Evaluation of Erasure Effects on Six Exposed Body Parts}
\label{app:six_part}

\autoref{fig:bar_rab} further demonstrates the defense effectiveness of the baseline methods and \name on the RAB adversarial prompt dataset. We use NudeNet\cite{nudenet} to detect exposed content in the generated images, which can identify various categories of exposure, such as female breasts (F\_Breast) and male genitalia (M\_Genitalia). It is noteworthy that the RAB attack generates more explicit content compared to I2P by adding extra pornographic semantics to the target prompt (the number of exposed content increases from 674 to 3596). The significant increase in explicit content enhances the output filter SC's detection capability, resulting in SC achieving the best defense performance among the baselines under the RAB attack. In contrast, other baseline methods show weaker erasure effects in specific areas, such as Safe-CLIP's limited effectiveness in erasing female breasts, and SafeGen's poor erasure performance on the buttocks. However, \name effectively suppresses the generation of all exposed body parts, with an ER of 99.14\% for total exposed areas, achieving SOTA performance.

% \section{Ethical Considerations} 
% To handle sensitive content responsibly, we follow previous work\cite{li2024safegen, gandikota2023erasing} by using automated methods\cite{laion-nsfw-detector, schramowski2022can} to evaluate generated content, minimizing manual review and exposure to harmful material.
% In addition, for inappropriate content displayed in the manuscript, we apply masking measures to ensure that readers are not directly exposed to sensitive materials, thus adhering to ethical standards for handling such content.

\end{document}